# Motion Planning for Multiple Unit-Ball Robots in $\mathbb{R}^d$

Israela Solomon and Dan Halperin

Blavatnik School of Computer Science, Tel-Aviv University, Israel

**Abstract.** We present a decoupled algorithm for motion planning for a collection of unit-balls moving among polyhedral obstacles in $\mathbb{R}^d$, for any $d \geqslant 2$. We assume that the robots have *revolving areas* in the vicinity of their start and target positions: Revolving areas are regions where robots can maneuver in order to give way to another moving robot. Given that this assumption is fulfilled, the algorithm is complete, namely it is guaranteed to find a solution or report that none exists. A key goal in our design is to make the revolving areas as economical as possible and in particular to allow different revolving areas to overlap. This makes the analysis rather involved but in return makes the algorithm conceptually fairly simple. We show that for the case of $m$ unit-discs moving among polygonal obstacles of total complexity $n$ in $\mathbb{R}^2$, the algorithm can be executed in $O(n^2m + m(m+n)\log(m+n))$ time. We implemented the algorithm for this case and tested it on several scenarios, for which we show experimental results for up to 1000 robots. Finally, we address the problem of choosing the order of execution of the paths in decoupled algorithms that locally solve interferences and show that finding the optimal order of execution is NP-hard. This motivated us to develop a heuristic for choosing the order; we describe the heuristic and demonstrate its effectiveness in certain scenarios.

## 1 Introduction

The basic motion planning problem in robotics is to produce a continuous path for a robot that connects a given start position and a given target position, while avoiding collisions with given obstacles. The geometric description of the robot and the obstacles is known, and the execution of the path is exact.

A natural extension of this problem is *multi-robot motion planning*, in which multiple robots share a common workspace. In this variant, each robot has its own start and target positions, and the goal is to produce a path for each robot with the additional requirement that the paths do not incur collisions between robots.

### 1.1 Previous Work

The first algorithms for multi-robot motion planning were *complete* algorithms, guaranteed to find a solution when one exists or report that none exists otherwise. In the seminal "piano movers" series of papers, Schwartz and Sharir gave the first algorithm for moving multiple general bodies in the plane [26], and another specialized algorithm for moving discs in the plane [27]. Their algorithms are polynomial in the combinatorial complexity of the obstacles, which we denote by $n$, but exponential in the number of robots (in particular, their running time is $O(n^3)$ for two discs and $O(n^{13})$ for three discs). Later on, Yap [38] gave algorithms for two and three discs whose running times are $O(n^2)$ and $O(n^3)$ respectively. Shortly after,

Sifrony and Sharir [29] gave another $O(n^2)$ algorithm for two convex translating robots, each of constant descriptive complexity, which is extensible to $m$ robots in $O(n^m)$ time.

It has been shown that several elementary variants of the multi-robot problem are computationally intractable. Hopcroft et al. [14] proved that the problem is PSPACE-hard for rectangular robots bounded in a rectangular region. Their result was later improved by Hearn and Demaine [12] to rectangles whose sizes are restricted to 1x2 and 2x1. Spirakis and Yap [33] proved that the problem is NP-hard for disc robots of varying radii in a simple polygon workspace.

Due to the hardness results, most of the subsequent research was focused on developing heuristic and approximation algorithms. Specifically, the most common approach since then has been *sampling-based algorithms* [6, Chapter 7],[19]. Sampling-based algorithms try to capture the structure of the configuration space without explicitly constructing it, by sampling random configurations and constructing a roadmap based on these configurations, in which a path can be looked for. These algorithms often provide asymptotic guarantees of completeness and optimality. However, they can only find paths in reasonable time if the paths do not pass through narrow or tight passages (since the probability to sample the points along such passages may become minuscule), they are incapable of determining whether a solution does not exist, and their running time is unbounded.

Approaches for multi-robot motion planning can be generally classified as *centralized* or *decoupled*. Centralized approaches consider all the robots as a single composite robot and compute a path in the combined configuration space, whose dimension is the sum of the number of degrees of freedom of the individual robots. Most of the motion planning techniques for a single robot can be applied as-is to the composite robot. When dealing with a large number of robots, the combined configuration space is of high dimension, thus the centralized techniques tend to be slow. In contrast, decoupled techniques find a path for each robot individually (or for small subsets of the robots) and then combine the paths into a joint solution while avoiding collisions between the robots. Approaches for avoiding collisions include adjusting the velocities of the robots (e.g. [7,15,20,21,22,24]), resolving collisions by local modifications of the paths (e.g. [11]), and prioritizing the robots and treating robots with higher priority as obstacles (e.g. [2,5,8,9,25]). Decoupled algorithms are usually faster (though some of them are heuristics whose running time is not theoretically bounded). But, unless restricting the input, they are usually not complete, as coupling is sometimes necessary, for example when the target position of each robot is the start position of some other robot.

Another variant of the multi-robot problem is the *unlabeled* multi-robot motion planning problem, in which the robots are indistinguishable and any target position can be covered by any robot. Similarly to the labeled problem, the unlabeled problem was also shown to be computationally intractable; Solovey and Halperin [31] proved it to be PSPACE-hard for unit-square robots amidst polygonal obstacles. A couple of works presented complete polynomial-time algorithms for unlabeled disc robots under simplifying assumptions, that the start and target positions have some minimal distance from each other and from the obstacles ([1,32,35]). Some of these works provide optimality guarantees: The solution by Turpin et al. [35] is optimal with respect to the length of the longest path, whereas the solution by Solovey et al. [32] is near-optimal with respect to the total length of all the paths.

## 1.2 Contribution

We present a decoupled algorithm for motion planning for a collection of unit-balls moving among polyhedral obstacles in $\mathbb{R}^d$, for any $d \geqslant 2$.

Our algorithm makes a simplifying assumption regarding the input. It requires that for each start or target position $z$ there exists a ball of radius 2, termed *revolving area*, that contains the unit-ball centered at $z$, and does not intersect any obstacle or any unit-ball centered at any other start or target position. Given that this assumption is fulfilled, our algorithm is *complete*, namely, it is guaranteed to find a solution if one exists, or report that none exists otherwise.

Following the methodology of decoupled algorithms, our algorithm first finds a path for each robot, disregarding other robots, and then modifies the paths to avoid collision between the robots. The paths are planned to be executed one after the other. Our simplifying assumption allows the algorithm to temporarily move the robots in the vicinities of their start and target positions, thus avoiding collisions between the robots while making only local modifications to the paths.

The algorithm uses several operations as black boxes, for which we suggest a few possible implementations. The running time of the algorithm, as well as the total length and the total combinatorial complexity of the resulting paths, depend on the specific methods used to perform these operations. We show that for the case of $m$ unit-discs moving among polygonal obstacles of total combinatorial complexity $n$ in $\mathbb{R}^2$, the algorithm can be executed in $O(n^2m + m(m+n)\log(m+n))$ time, when using shortest paths as the initial paths.

The algorithm has been implemented in code for unit-discs moving among polygonal obstacles in the plane. We tested our implementation on several scenarios, for which we show experimental results for up to $1000$ robots. The results attest to the efficiency of the algorithm and suggest that the excess total length of the paths over the original paths is small in practice (although we show a theoretical example in which the excess length tends to infinity, as the number of obstacles tends to infinity and their size tends to zero).

Finally, we address the problem of choosing the order of execution of the paths. Since our algorithm first finds an initial path for each robot and then locally handles interferences, the order in which the paths are executed might have a substantial effect on the running time and on the properties of the paths that the algorithm finds. Thus, it is desirable to choose an order that minimizes the number of interferences that the algorithm has to handle. We formalize this problem more generally, for any decoupled algorithm that locally handles interferences, and prove that it is NP-hard to find the optimal order. We also describe a heuristic for choosing the order and demonstrate its effectiveness in certain scenarios.

**Remark** Without any simplifying assumptions, the multi-robot motion-planning problem is intractable (see the Previous Work section above). The best known general algorithms have exponential dependence on the total number of degrees of freedom of all the robots. To obtain efficient solutions we need to make some assumptions. The success of sampling-based planners has taught us that clearance in the workspace (the availability of sufficient distance to the obstacles and distance between the moving robots) is a key factor to efficiency in practice. We argue that our revolving-areas assumption keeps the problem non-trivial still. Notice that we do not assume anything about the clearance of the paths of the robots but rather only assume that the robots in their start and goal positions have the required revolving areas. Furthermore, our assumptions do *not* preclude that a pair of robots will be osculating at their start or target positions, as long as otherwise each of them has a revolving area not containing the other robot. Similarly, we do not preclude start or goal positions with zero clearance from the obstacles. Additionally, we allow revolving areas to overlap. All these together make the problem challenging to analyze and solve.

## 2 Algorithmic Framework

### 2.1 Preliminaries and Assumptions

We consider the problem of $m$ unit-ball robots moving in a polyhedral workspace $\mathcal{W} \subseteq \mathbb{R}^d$. We assume that the dimension $d$ is fixed. We define $\mathcal{O} := \mathbb{R}^d \setminus \mathcal{W}$ to be the complement of the workspace, and we call $\mathcal{O}$ the obstacle space. We denote by $n$ the complexity of the workspace, which is the total number of faces of any dimension of the obstacles.

For every $x \in \mathbb{R}^d$, let $\mathcal{D}_x$ be the open unit-ball centered at $x$. The set of positions in which a unit-ball robot does not collide with the obstacle space, termed the free space, is $\mathcal{F} := \{x \in \mathcal{W} : \mathcal{D}_x \cap \mathcal{O} = \emptyset\}$. Notice that we denote a position of a ball robot by the coordinates of its center. A path is a continuous function $\gamma : [0,1] \to \mathbb{R}^d$. We say that a path is collision-free if its image is contained in $\mathcal{F}$. For a set $X \subseteq \mathbb{R}^d$, we denote its boundary by $\partial X$.

Given a set of start positions $S = \{s_1, ..., s_m\}$ and a set of target positions $T = \{t_1, ..., t_m\}$, our goal is to plan a collision-free motion for the $m$ unit-ball robots, such that for every $1 \leqslant i \leqslant m$ there is one path that starts at $s_i$ and ends at $t_i$. Formally, for each $1 \leqslant i \leqslant m$ we wish to find a path $\gamma_i : [0,1] \to \mathcal{F}$, such that $\gamma_i(0) = s_i$ and $\gamma_i(1) = t_i$. For every $i \neq j$ and $t \in [0,1]$, we require that $\mathcal{D}_{\gamma_i(t)} \cap \mathcal{D}_{\gamma_j(t)} = \emptyset$, so the robots will not collide with each other.

We assume that each start or target position $z$ has a *revolving area* in which a robot placed at $z$ can move in order to avoid collision with any other robot. Formally, we assume that for every $z$, $\mathcal{D}_z$ can be placed inside a (not necessarily concentric) open ball $\mathcal{A}_z$ of radius 2 such that:

- $\mathcal{A}_z \cap \mathcal{O} = \emptyset$; that is, $\mathcal{A}_z$ does not intersect any obstacle.
- For every start or target position $y \neq z$, $\mathcal{A}_z \cap \mathcal{D}_y = \emptyset$.

We call $\mathcal{A}_z$ the revolving area of $z$, and denote its center by $c_z$. Note that the revolving areas may intersect each other. We also let $\mathcal{C}_z := \mathcal{D}_{c_z}$ and denote by $\mathcal{B}_z$ the ball of radius 3 centered at $c_z$. See Figure 1.

Allowing the revolving areas to intersect makes the method less restrictive and applicable to a larger set of scenarios than if we had assumed that they do not intersect. Much of the effort in the forthcoming analysis goes to accommodating this requirement.

Throughout the paper we use the following useful lemma.

**Lemma 1.** *For any two revolving areas, the distance between their centers is at least* 2.

*Proof.* Let $s_j, s_k$ be two start or target positions. Since $\mathcal{D}_{s_j} \cap \mathcal{A}_{s_k} = \emptyset$ and the radius of $\mathcal{A}_{s_k}$ is 2, then for any $p \in \mathcal{D}_{s_j}$ it holds that $\|p - c_{s_k}\| \geqslant 2$. Noting that $c_{s_j}$ belongs to the closure of $\mathcal{D}_{s_j}$, it follows that $\|c_{s_j} - c_{s_k}\| \geqslant 2$.

Lemma 1 implies that any ball of constant radius contains only $O(1)$ start and target positions.

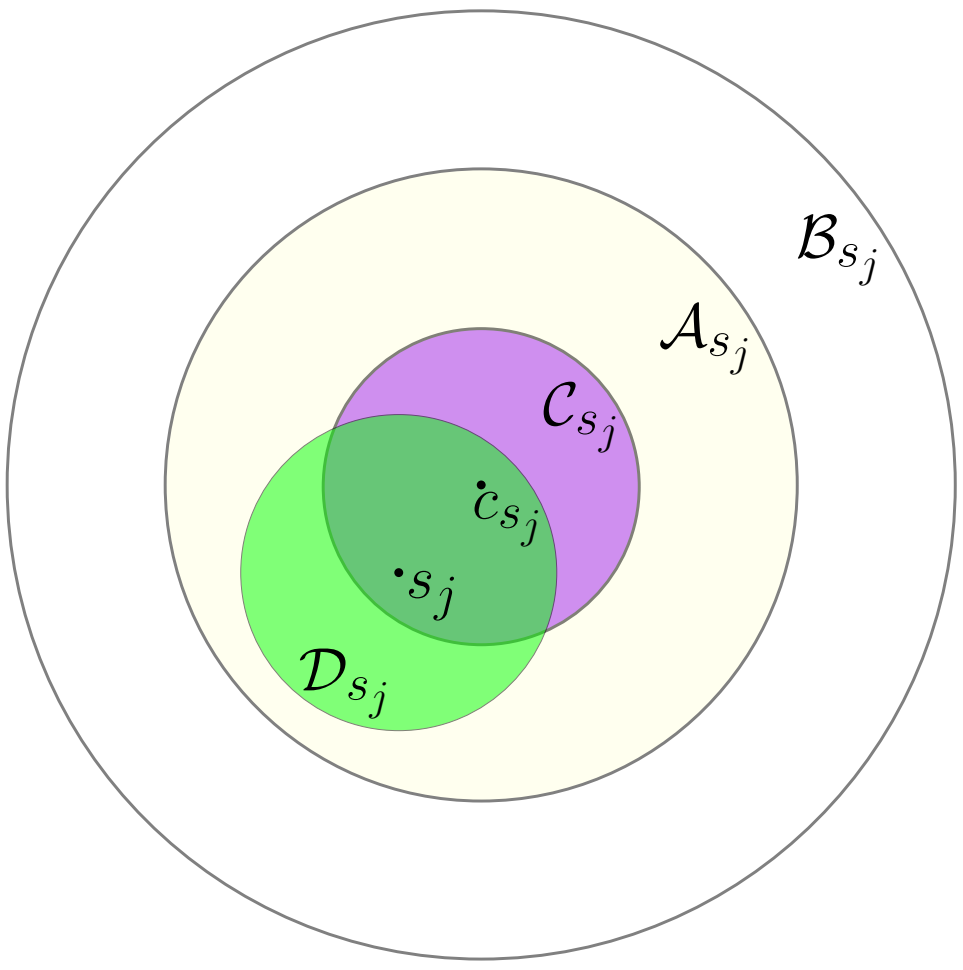

**Fig. 1:** The ball $\mathcal{A}_{s_j}$ is the revolving area of the start position $s_j$. The unit-ball $\mathcal{D}_{s_j}$ is centered at $s_j$ and is contained in the revolving area $\mathcal{A}_{s_j}$. The unit-ball $\mathcal{C}_{s_j}$ and the ball $\mathcal{B}_{s_j}$ of radius 3 are centered at $c_{s_j}$, which is the center of $\mathcal{A}_{s_j}$.

### 2.2 Algorithm Overview

Our algorithm operates as follows. First, it finds a set of free paths $\{\gamma_1, \ldots, \gamma_m\}$, where $\gamma_i : [0,1] \to \mathcal{F}$ is the path of robot $r_i$ from $s_i$ to $t_i$, ignoring possible collisions with other robots. Then, it modifies these paths in order to avoid collisions between the robots. The algorithm iterates over the paths in an arbitrary order, and based on each path $\gamma_i$, it builds a new path $\gamma'_i$ from $s_i$ to $t_i$ that does not pass "too close" to centers of revolving areas. When a robot follows the new path, other robots can be moved inside their revolving areas according to *retraction paths* and avoid collisions.

In Section 2.3 we define the retraction paths and show that they can be used to avoid collisions. In Section 2.4 we show how to build the new paths. In Section 2.5 we describe the algorithm in detail. In Section 3 we suggest possible concrete implementations to operations the algorithm uses as "black boxes". In Section 4 we analyze the algorithm.

### 2.3 Retraction

In this section we are concerned with moving a single robot $r_i$ along its path $\gamma_i$. We assume that the other robots are positioned at their start or target positions. Without loss of generality, assume that every $r_j$, for $j \neq i$, is positioned at its start position $s_j$. During the motion of $r_i$ along $\gamma_i$ it might interfere (collide) with other robots. We wish to show that such interferences can be avoided by slightly modifying $\gamma_i$ and temporarily moving the interfering robots inside the respective revolving areas $\mathcal{A}_{s_j}$.

More formally, suppose there exist $x \in \gamma_i$ and $1 \leqslant j \neq i \leqslant m$ such that $\mathcal{D}_x \cap \mathcal{A}_{s_j} \neq \emptyset$ and $x \notin \mathcal{C}_{s_j}$. We show that there exists a point $\rho_{s_j}(x) \in \mathcal{A}_{s_j}$ to which $r_j$ can be moved so that $r_i$ will not collide with $r_j$, where $r_i$ is at $x$ and $r_j$ is at $\rho_{s_j}(x)$. The *retraction point* $\rho_{s_j}(x)$ is defined so that several robots that are *retracted* by $x$ do not collide.

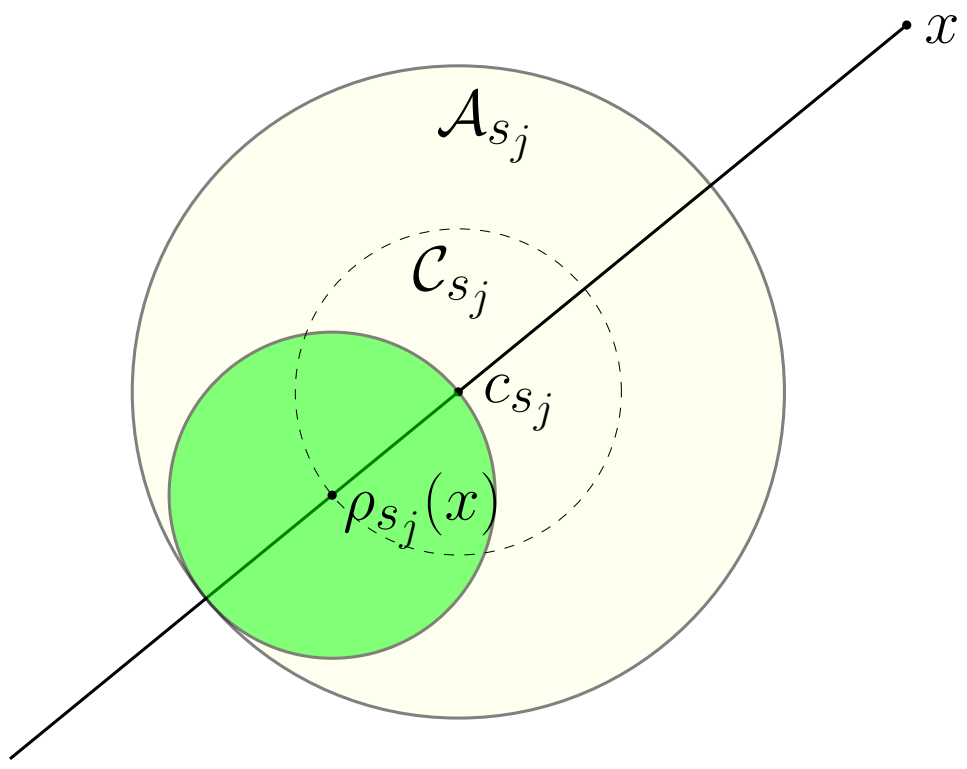

**Fig. 2:** The retraction point from $s_j$ with respect to $x$.

**Definition 1.** *Let $x$ be a point in $\mathcal{W}$ and assume that $x \neq c_{s_j}$, for some $s_j$. The* retraction point *from $s_j$ with respect to $x$ is the unique point $\rho_{s_j}(x)$ on the line through $c_{s_j}$ and $x$, such that (1) $\mathcal{D}_{\rho_{s_j}(x)} \subseteq \mathcal{A}_{s_j}$; (2) $\mathcal{D}_{\rho_{s_j}(x)}$ touches $\partial\mathcal{A}_{s_j}$; and (3) $c_{s_j}$ is between $x$ and $\rho_{s_j}(x)$ (see Figure 2).*

The following lemma states that, for any $i \neq j$, robot $r_i$ placed at $x$ does not collide with robot $r_j$ placed at the retraction point from some $s_j$ with respect to $x$.

**Lemma 2.** *If $x \notin \mathcal{C}_{s_j}$, then $\mathcal{D}_x \cap \mathcal{D}_{\rho_{s_j}(x)} = \emptyset$.*

*Proof.* $\|x - \rho_{s_j}(x)\| = \|x - c_{s_j}\| + \|c_{s_j} - \rho_{s_j}(x)\| \geqslant 1 + 1 = 2$.

Next, we show that if multiple robots are retracted with respect to $x$, then, even though their respective revolving areas may intersect, they do not collide with each other.

**Lemma 3.** *For any $j \neq k$, and any point $x \neq c_{s_j}, c_{s_k}$, it holds that $\mathcal{D}_{\rho_{s_j}(x)} \cap \mathcal{D}_{\rho_{s_k}(x)} = \emptyset$.*

*Proof.* Let $r_j, r_k$ be two robots that are retracted with respect to $x$. By Lemma 1, the distance between the centers of the revolving areas of the robots is at least 2. Therefore, it is sufficient to prove that the distance between the retraction points $\rho_{s_j}(x)$ and $\rho_{s_k}(x)$ is greater than the distance between the centers of their revolving areas, which yields $\|\rho_{s_j}(x) - \rho_{s_k}(x)\| \geqslant \|c_{s_j} - c_{s_k}\| \geqslant 2$ as desired.

By the way $\rho_{s_j}(x)$ is defined, it follows that $\rho_{s_j}(x)$ is the unique point lying on the line through $x$ and $c_{s_j}$, at distance 1 from $c_{s_j}$, such that $c_{s_j}$ lies between $x$ and $\rho_{s_j}(x)$. Letting $a := \|x - c_{s_j}\|$, we have $\|x - \rho_{s_j}(x)\| = a + 1$. Similarly, for $b := \|x - c_{s_k}\|$, it holds that $\|x - \rho_{s_k}(x)\| = b + 1$, and that $c_{s_k}$ lies between $x$ and $\rho_{s_k}(x)$. We use the law of cosines. Denote by $\alpha$ the angle between the line through $x$, $c_{s_j}$ and $\rho_{s_j}(x)$ and the line through $x$, $c_{s_k}$

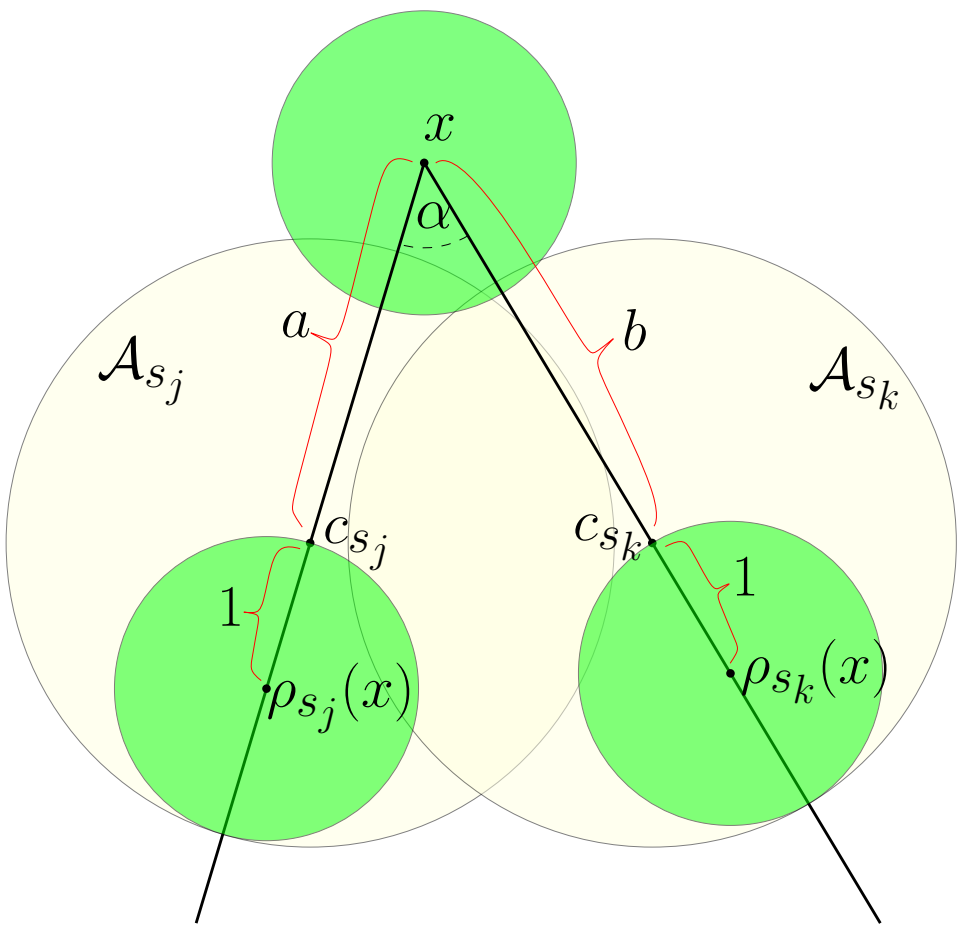

**Fig. 3:** Illustration of the proof of Lemma 3.

and $\rho_{s_k}(x)$. See Figure 3.

$$\begin{aligned}\|\rho_{s_j}(x) - \rho_{s_k}(x)\|^2 - &\|c_{s_j} - c_{s_k}\|^2 \\ &= \big((a+1)^2 + (b+1)^2 - 2(a+1)(b+1)\cos\alpha\big) - \big(a^2 + b^2 - 2ab\cos\alpha\big) \\ &= 2a + 2b + 2 - 2(a+b+1)\cos\alpha \\ &= (2a+2b+2)(1-\cos\alpha) \\ &\geqslant 0.\end{aligned}$$

Rearranging the equation, we conclude that $\|\rho_{s_j}(x) - \rho_{s_k}(x)\| \geqslant \|c_{s_j} - c_{s_k}\| \geqslant 2$.

We extend the definition of a retraction from a single point to a path.

**Definition 2.** *Let $\gamma : I \subseteq [0,1] \to \mathcal{F}$ be a path that does not pass through $c_{s_j}$ for some $s_j$. The* retraction *from $s_j$ with respect to $\gamma$ is the path $\mathcal{P}_{\gamma,s_j} : I \to \mathcal{A}_{s_j}$ defined by $\mathcal{P}_{\gamma,s_j}(t) := \rho_{s_j}(\gamma(t))$ for all $t \in I$.*

Applying Lemma 2 and Lemma 3 on each point in a retraction path, we conclude that if $r_i$ follows $\gamma$ and $r_j$ and $r_k$ follow their retraction paths with respect to $\gamma$, they do not collide with each other or with $r_i$.

**Corollary 1.** *Let $\gamma : I \subseteq [0,1] \to \mathcal{F}$ be a path for robot $r_i$ that does not pass through $\mathcal{C}_{s_j}$ for all $j \neq i$. Then for any $t \in I$ and any $1 \leqslant j \neq k \leqslant m$, it holds that $\|\gamma(t) - \mathcal{P}_{\gamma,s_j}(t)\| \geqslant 2$ and that $\|\mathcal{P}_{\gamma,s_j}(t) - \mathcal{P}_{\gamma,s_k}(t)\| \geqslant 2$.*

Next we show that a robot $r_j$ can be moved from its start position to the beginning of its retraction path with respect to $x$ on the segment between them, where $\mathcal{D}_x$ touches but does not intersect the revolving area containing $r_j$. Since during this motion $r_j$ is contained in its revolving area, it cannot collide with $r_i$ or with other robots that are positioned at their start or target positions. We show that this motion also does not cause collision with another robot that is retracted with respect to $x$.

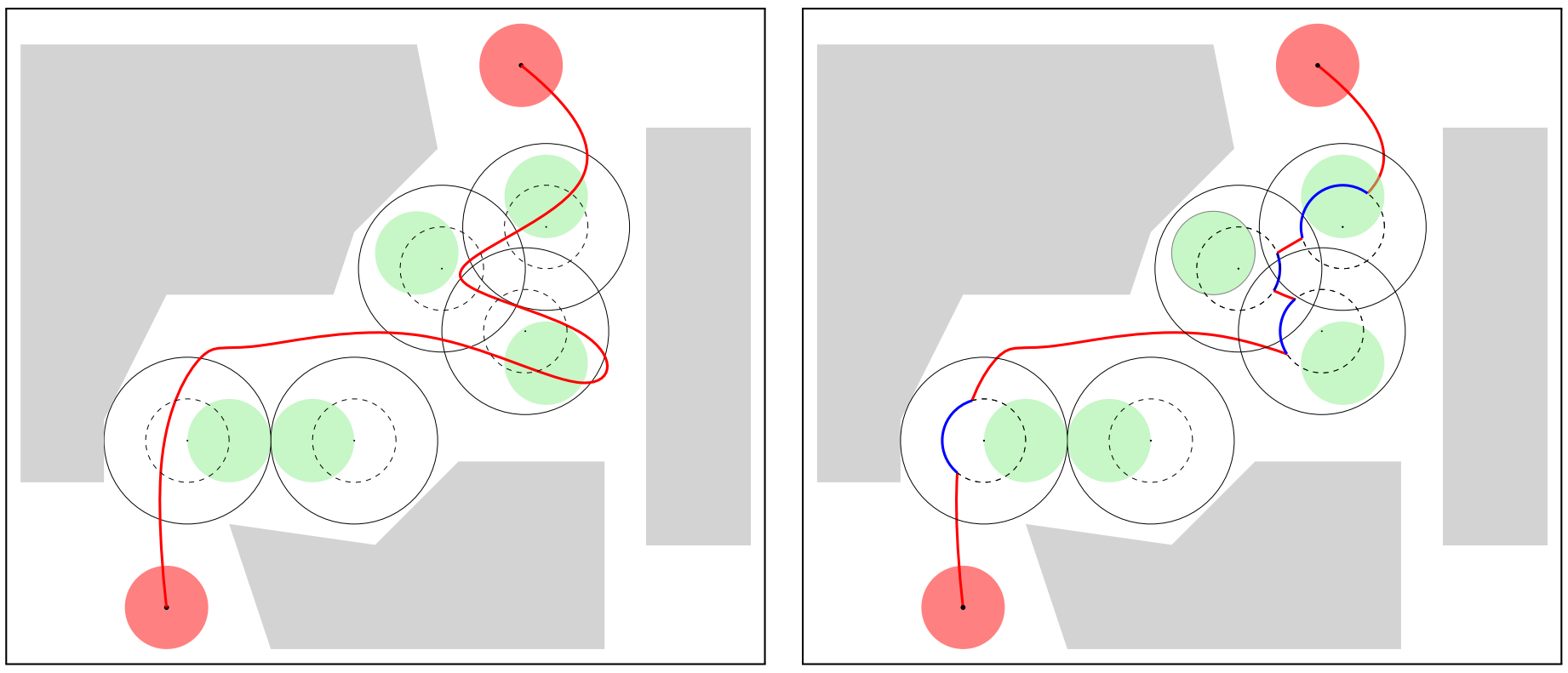

**(a)** original path **(b)** new path

**Fig. 4**

The following proposition states that if $\mathcal{D}_x$ touches the revolving area of a robot $r_j$ and touches or intersects the revolving area of another robot $r_k$, which is retracted with respect to $x$, then $r_j$ can be moved inside its revolving area to the retraction point from $s_j$ with respect to $x$ without colliding with $r_k$.

**Proposition 1.** *Let $x \in \mathcal{W}$ such that $\|x - c_{s_j}\| = 3$ and $1 \leqslant \|x - c_{s_k}\| \leqslant 3$. Then for any point $u$ in the segment $[\rho_{s_j}(x), s_j]$, it holds that $\|\rho_{s_k}(x) - u\| \geqslant 2$.*

The proof appears in the appendix.

### 2.4 Constructing a New Path

In this section we show how to modify a path, so a robot that follows the modified path will not collide with any robot that moves along the retraction paths defined above.

Assume we are given a path $\gamma : [0,1] \to \mathcal{F}$ from $s_i$ to $t_i$ for a robot $r_i$. We assume that all the other robots are in their start or target positions, and denote the set of start and target positions that are currently occupied by robots other than $r_i$ by $Z \subseteq S \cup T$. We show how to construct a new path $\gamma' : [0,1] \to \mathcal{F}$ from $s_i$ to $t_i$, such that $\gamma'$ does not pass through $\mathcal{C}_z$ for any $z \in Z$. Notice that this path may still collide with other robots, but this will be resolved later by moving the robots along retraction paths.

For any $z \in Z$ and an intersection point $p = \gamma(t)$ of $\gamma$ and $\partial\mathcal{C}_z$, we call $p$ an *entrance point to* $\mathcal{C}_z$ if $\gamma(t+\varepsilon) \in \mathcal{C}_z$ and $\gamma(t-\varepsilon) \notin \mathcal{C}_z$, or an *exit point from* $\mathcal{C}_z$ if $\gamma(t-\varepsilon) \in \mathcal{C}_z$ and $\gamma(t+\varepsilon) \notin \mathcal{C}_z$, for infinitesimally small $\varepsilon > 0$.

First the algorithm finds the intersection points of $\gamma$ with every $\partial\mathcal{C}_z$ for $z \in Z$ that are either entrance points or exit points, and sorts all these points according to their order of appearance in $\gamma$. The algorithm starts constructing a new path $\gamma'$ from $s := s_i$. Note that $s \notin \mathcal{C}_z$ for all $z \in Z$, and this invariant holds during the algorithm for every value that is assigned to $s$.

Then the algorithm repeats the following recursively. If the whole sub-path of $\gamma$ from $s$ to $t_i$ does not intersect any $\mathcal{C}_z$ for $z \in Z$, then this sub-path is added to $\gamma'$, and the algorithm terminates. Otherwise, the algorithm finds the first entrance point to some $\mathcal{C}_z$. Denote this entrance point by $p_z$, and denote the last exit point from $\mathcal{C}_z$ by $q_z$. Then the algorithm adds to $\gamma'$ the sub-path of $\gamma$ from $s$ to $p_z$ and a geodesic arc of $\partial\mathcal{C}_z$ from $p_z$ to $q_z$. Afterwards, the algorithm sets $s := q_z$ and recursively repeats the process.

**Lemma 4.** *The new path $\gamma'$ is free from collision with the obstacles and does not intersect $\mathcal{C}_z$ for any $z \in Z$.*

*Proof.* The path $\gamma'$ is composed of two types of sub-paths: (1) sub-paths of $\gamma$ that are free and do not intersect any such $\mathcal{C}_z$; and (2) geodesic arcs of $\partial\mathcal{C}_z$ for $z \in Z$. Since $\mathcal{C}_z$ is a unit-ball centered at the center of a revolving area, which is a ball of radius 2 contained in the workspace $\mathcal{W}$, it follows that $\partial\mathcal{C}_z \subseteq \mathcal{F}$. Moreover, by Lemma 1, for every $z \neq y \in Z$ it holds that $\mathcal{C}_z \cap \mathcal{C}_y = \emptyset$. Therefore an arc of $\partial\mathcal{C}_z$ does not intersect any $\mathcal{C}_y$.

### 2.5 The Algorithm

We now describe our algorithm in detail. First, we find a set of paths $\{\gamma_1, \ldots, \gamma_m\}$, where $\gamma_i : [0,1] \to \mathcal{F}$ is the path of robot $r_i$ from $s_i$ to $t_i$, avoiding the polyhedral obstacles, but ignoring the other robots. The algorithm iterates over the paths, and for each $i$ it builds a new path $\gamma'_i$ for robot $r_i$ while assuming that $r_j$ is in its start position for all $j > i$ or in its target position for all $j < i$. We denote by $Z_i := \{t_j \mid 1 \leqslant j < i\} \cup \{s_j \mid i < j \leqslant m\}$ the set of start and target positions that are occupied during the motion of $r_i$. The new path $\gamma'_i$ from $s_i$ to $t_i$ is built as described in Section 2.4. By Lemma 4, $\gamma'_i$ is a path that is free from collision with the obstacles and does not intersect $\mathcal{C}_z$ for any $z \in Z_i$.

We find a reparametrization of $\gamma'_i$ such that whenever $r_i$ follows the path and is about to enter (or leave) a revolving area containing another robot, $r_i$ stays at its position while the other robot moves to (or from) its retraction path. Formally, for each path $\gamma'_i$, the algorithm finds a reparametrization $\gamma''_i = \gamma'_i \circ f_i$, where $f_i : [0,1] \to [0,1]$ is a continuous non-decreasing function, which satisfies the following. Consider any $\zeta \in [0,1]$ and any revolving area $\mathcal{A}_z$, $z \in Z_i$, such that $\|\gamma'_i(\zeta) - z\| = 3$ and $\|\gamma'_i(\zeta+\varepsilon) - z\| < 3$ or $\|\gamma'_i(\zeta-\varepsilon) - z\| < 3$ for infinitesimally small $\varepsilon > 0$. Then $f_i^{-1}(\zeta)$ is a segment of the form $[\zeta', \zeta' + \delta]$, where $\delta > 0$ is a fixed small constant. It means that as $r_i$ follows $\gamma''_i$, whenever $r_i$ touches $\partial\mathcal{A}_z$ and is about to intersect $\mathcal{A}_z$, or just stopped intersecting it, there is a time frame of length $\delta$ in which $r_i$ stays at the same position.

We now define the final paths $\pi_1, \ldots, \pi_m : [0,m] \to \mathcal{F}$. At first, for each $i$ we set $\pi_i([0, i-1]) := s_i$, $\pi_i([i-1, i]) := \gamma''_i$ and $\pi_i([i,m]) := t_i$. Next, retraction paths are added for sub-paths that might incur collision. Roughly, whenever a robot following the path $\pi_i$ is about to enter a revolving area $\mathcal{A}_z$ that contains a robot $r_j$, the robot $r_j$ is moved to the retraction point from $z$ with respect to the position of the robot $r_i$. Then $r_j$ follows the retraction path $\mathcal{P}_{\pi_i,z}$ as long as $r_i$ intersects $\mathcal{A}_z$. When $r_i$ leaves $\mathcal{A}_z$, $r_j$ is moved back to $z$, its start or target position.

Formally, consider any $1 \leqslant i \neq j \leqslant m$ and a maximal time segment $(a,b) \subset [i-1, i]$ in which for all $\zeta \in (a,b)$ it holds that $\mathcal{D}_{\pi_i(\zeta)}$ intersects $\mathcal{A}_z$, where $\mathcal{A}_z$ is the revolving area that contains the robot $r_j$ at time $[i-1, i]$. Because of the reparametrization, it holds that $\pi_i(\zeta) = \pi_i(a)$ for any $\zeta \in [a-\delta, a]$ and that $\pi_i(\zeta) = \pi_i(b)$ for any $\zeta \in [b, b+\delta]$. We change $\pi_j$ to follow the retraction path $\mathcal{P}_{\pi_i,z}$ in this time frame. We set $\pi_j((a,b)) := \mathcal{P}_{\pi_i,z}((a,b))$, and we set $\pi_j([a-\delta, a])$ and $\pi_j([b, b+\delta])$ to be the segments from $z$ to $\mathcal{P}_{\pi_i,z}(a)$ and from $\mathcal{P}_{\pi_i,z}(b)$ to $z$, respectively.

In case that as $r_i$ follows $\gamma_i'$ it touches and is about to intersect (or stop intersecting) several occupied revolving areas at the same time $\zeta$, then the reparametrization is defined such that $f_i^{-1}(\zeta)$ is a time segment $\hat{\delta}$ whose length is $\delta$ times the number of these revolving areas. Then, we add retraction paths for all those revolving areas at consecutive time segments of length $\delta$ during the time segment $\hat{\delta}$. It means that the robots are moved to the beginning (or from the end) of their retraction paths one after the other, while $r_i$ stays at its position.

Putting it all together, by Corollary 1 and Proposition 1 the paths $\pi_1, \ldots, \pi_m$ are free and do not incur collision between the robots.

## 3 Concrete Methods

There are several operations that we use in the algorithm as "black boxes": finding revolving areas, finding the original paths for the robots and computing intersections between paths and spheres. These operations can be implemented in different ways, resulting in different running time and different paths. We suggest a few methods to perform these steps.

### 3.1 Finding Revolving Areas

***General Method*** A point can serve as the center of the revolving area of a start or target position $z$ if and only if its distance from $z$ is at most 1 and its distance from $\mathcal{O} \cup \{\mathcal{D}_y\}_{z \neq y \in S \cup T}$ is at least 2. Therefore, for any start or target position $z$ we can find a revolving area (or conclude that no such ball exists) in the following manner. Compute the union of the Minkowski sums of a ball of radius 2 with each obstacle and with each unit-ball $\mathcal{D}_y$ centered at a start or target position $y \neq z$. Denote the complement of this union by $\mathcal{N}$. Find a point in the intersection of $\mathcal{D}_z$ and $\mathcal{N}$. This point can be the center of the revolving area of $z$. If $\mathcal{D}_z \cap \mathcal{N}$ is empty, then no such ball exists.

Moreover, notice that it is sufficient to consider only start and target positions $y \neq z$ such that $\|y - z\| \leqslant 4$, since otherwise the Minkowski sum of $D_y$ and a ball of radius 2 does not intersect $D_z$. There are only $O(1)$ such start and target positions (by Lemma 1), and we denote the set of these points by $\mathcal{RB}(z)$.

We work out the details for moving discs in $\mathbb{R}^2$ and moving balls in $\mathbb{R}^3$. Recall that $n$ is the complexity of the polyhedral workspace. In the following lemma we use the notation $\lambda_q(n)$ to denote the near-linear function related to Davenport-Schinzel sequences as defined in [28].

**Lemma 5.** *After a preprocessing phase whose time complexity is $O(m + n\log^2(n))$ in $\mathbb{R}^2$ and $O(m + n^2\lambda_q(n)\log(n))$ in $\mathbb{R}^3$, finding a center for a single revolving area (or reporting that none exists) can be done in $O(n)$ time in $\mathbb{R}^2$ and in $O(n^2\lambda_q(n))$ time in $\mathbb{R}^3$, where $q$ is a small constant. Therefore finding revolving areas for all the $2m$ start and target positions can be done in $O(n\log^2(n) + mn)$ time in $\mathbb{R}^2$ and in $O(n^2\lambda_q(n)\log(n) + mn^2\lambda_q(n))$ time in $\mathbb{R}^3$.*

*Proof.* The preprocessing stage is composed of two tasks: (i) finding the set $\mathcal{RB}(z)$ for each start or target position $z$, and (ii) processing the obstacles and their complement into a set of simple cells, each of complexity $O(1)$, that will be used when looking for revolving areas.

In order to perform (i) we tile $\mathbb{R}^d$ with a $d$-dimensional grid of unit cubes.[1] We construct a hash table that maps each cell of the grid to the set of start and target positions that are contained in this cell. We add each start or target position in $O(1)$ time to its cell in the hash table. By Lemma 1, each cell is mapped to $O(1)$ start and target positions. Then, for each start or target position $z$ we query the hash table for the cells that contain points whose distance from $z$ is at most 4. There are only $O(1)$ such cells and together they contain $O(1)$ start and target positions, among them $\mathcal{RB}(z)$. Therefore $\mathcal{RB}(z)$ can be obtained from them in $O(1)$ time. Thus the total running time of finding $\mathcal{RB}(z)$ for all $z$ is $O(m)$.

For (ii) we devise different methods for the cases of $\mathbb{R}^2$ and $\mathbb{R}^3$. We start with $\mathbb{R}^2$. We compute the union of the Minkowski sums of a disc of radius 2 with each obstacle in $O(n\log^2(n))$ time [17]. The combinatorial complexity of this union is $O(n)$ [17]. Then, we compute a vertical decomposition of $\mathcal{N}$, the complement of this union, in $O(n\log(n))$ time [4, Chapter 6]. This decomposition yields $O(n)$ cells, each of complexity $O(1)$, such that the union of their closures is exactly $\mathcal{N}$. The running time of this method is $O(n\log^2(n))$.

The method we devise for $\mathbb{R}^3$ is similar but more involved. We triangulate each two-dimensional face of the obstacles, so we get $O(n)$ triangles such that the union of their closures is the boundary of the obstacles. For each triangle, we compute its Minkowski sum with a ball of radius 2, and decompose the boundary of the result to $O(1)$ $xy$-monotone surfaces (which are patches of planes, balls and cylinders). These are $O(n)$ surfaces in total. We compute a vertical decomposition of the arrangement of these $O(n)$ surfaces in $O(n^2\lambda_q(n)\log(n))$ time, where $q$ is a constant [3]. This decomposition yields $O(n^2\lambda_q(n))$ cells, each of complexity $O(1)$. We mark the cells that are contained in the Minkowski sum of the triangles and a ball of radius 2 in $O(n^2\lambda_q(n))$ time. Consider the collection of the cells that are not marked. The union of the closures of these cells contains $\mathcal{N}$, and each cell is contained either in $\mathcal{N}$ or in $\mathcal{O}$. We let all of these unmarked cells be the output of (ii), so all of them will be used when applying the procedure described below for finding centers for revolving areas. Notice that applying this procedure to cells that are contained in $\mathcal{O}$ does not effect its output, since $\mathcal{O}$ and $\mathcal{D}_z$ are disjoint for any $z \in S \cup T$. The running time of this method is $O(n^2\lambda_q(n)\log(n))$.

After the preprocessing is done, finding a center for a revolving area for a start or target position $z$ can be done in the following manner. For each cell that was constructed in (ii), we compute the intersection of its closure with $\mathcal{D}_z$, and subtract from this intersection the balls of radius 3 centered at the start and target positions in $\mathcal{RB}(z)$ that were found in (i). If the result is a non-empty set, then any point in this set can serve as a center for a revolving area (otherwise, there is no such point in the closure of that cell). This computation involves $O(1)$ geometric objects, each of complexity $O(1)$, thus it takes $O(1)$ time for each cell. Therefore, after the preprocessing, finding a single revolving area takes $O(n)$ time in $\mathbb{R}^2$ and $O(n^2\lambda_q(n))$ time in $\mathbb{R}^3$.

***Stronger Assumptions*** In case it is known that each start or target position is at distance of at least 2 from any obstacle and at distance at least 3 from any other start or target position, then each start or target position can serve as the center of its revolving area.

[1] We assign the boundary to cells in some arbitrary consistent way.

### 3.2 Finding Original Paths

Any method for finding a single path for a ball robot moving among polyhedral obstacles can be used for producing the initial paths. We do not assume anything in particular about the original paths. These may as well be computed by sampling-based algorithms, especially in high dimensions. Notice however that the running time of the algorithm and the properties of the resulting paths depend on the properties of the initial paths, as explained in Section 4.

In Section 5 we analyze the case of unit-disc robots in $\mathbb{R}^2$ where the initial paths are each a shortest path for each robot while ignoring the other robots. In this case, a shortest path can be found in $O(n^2)$ time [13].

### 3.3 Computing Intersections

For each $1 \leqslant i \leqslant m$, we need to compute the intersection points of $\gamma_i$ with $\partial\mathcal{C}_z$ for all $z \in Z_i$, and the intersection points of $\gamma'_i$ with $\partial\mathcal{B}_z$ for all $z \in Z_i$. Recall that we are interested only in intersection points that are either entrance or exit points (as explained in Section 2.4), so even if an arc is contained in a sphere, we should consider only its endpoints.

Let $\gamma$ be a simple path and let $C$ be a set of $m$ spheres of constant radius, each concentric with a different revolving area. Denote by $k(\gamma)$ the combinatorial complexity of $\gamma$, namely, the number of arcs that $\gamma$ comprises, and by $I(\gamma, C)$ the number of intersection points between $\gamma$ and the spheres in $C$.

***Naive Approach*** Compute the intersection points of each arc of $\gamma$ and each sphere in $C$. The running time is $O(mk(\gamma))$.

***Sweep Line Algorithm in*** $\mathbb{R}^2$ We utilize the sweep line algorithm by Mairson and Stolfi [23] for detecting red-blue intersections between a set of red arcs and a set of blue arcs. The red arcs are the $O(k(\gamma))$ arcs of the path $\gamma$, whose relative interiors are pairwise disjoint, since the path is simple. The blue arcs are the arcs of the arrangement of the circles $C$. Note that the complexity of the arrangement is $O(m)$, since the distance between the centers of any two revolving areas is at least 2 (Lemma 1). Thus, there are $O(m)$ blue arcs, whose relative interiors are also pairwise disjoint. We split each arc (blue or red) into a finite number of $x$-monotone arcs. Therefore, we can find all the intersections of $\gamma$ and the circles in $C$ in $O((k(\gamma) + m)\log(k(\gamma) + m) + I(\gamma, C))$ time.

In Section 5 we show that for a shortest path $\gamma$ for a disc robot in $\mathbb{R}^2$ it holds that $I(\gamma, C) = O(k(\gamma) + m)$, and since the complexity of a shortest path for a disc robot in $\mathbb{R}^2$ among polygonal obstacles with a total number of $n$ vertices is $k(\gamma) = O(n)$, the running time of the method in this case is $O((m + n)\log(m + n))$.

## 4 General Analysis

As explained in Section 3, there are several steps in the algorithm that can be implemented by different methods, resulting in different running time and different paths. In this section we give a general analysis, in which we use some upper bounds that depend on the properties of the original paths and the algorithms that are used to perform these operations. Afterwards, in Section 5 we analyze the special case of shortest paths in $\mathbb{R}^2$.

As before, we denote by $k(\gamma)$ the complexity of $\gamma$. In addition we use the following upper bounds:

- $F(n,m)$ - The total time it takes to find for all $1 \leqslant i \leqslant m$ a free path $\gamma_i$ from $s_i$ to $t_i$, disregarding other robots.
- $M(n,m)$ - The total time it takes to find revolving areas for all $2m$ start and target positions.
- $K(n,m)$ - An upper bound on the total complexity of all the $m$ paths $\gamma_i$.
- $L(n,m)$ - An upper bound on the total length of all the $m$ paths $\gamma_i$.
- $Q(n,m)$ - The total time it takes to compute for all $1 \leqslant i \leqslant m$ the intersection points of $\gamma_i$ with $\partial\mathcal{C}_z$ and of $\gamma'_i$ with $\partial\mathcal{B}_z$ for all $z \in Z_i$.
- $I(n,m)$ - An upper bound on the total number of intersection points of $\gamma_i$ with $\partial\mathcal{C}_z$ and of $\gamma'_i$ with $\partial\mathcal{B}_z$, summed over all $1 \leqslant i \leqslant m$ and all $z \in Z_i$.

**Theorem 1.** *Given polyhedral workspace of complexity $n$ in $\mathbb{R}^d$, and for each of $m$ unit-ball robots a start position and a target position, the algorithm described above computes free paths for the robots in $O(M(n,m) + F(n,m) + Q(n,m) + I(n,m)\log(I(n,m)) + K(n,m) + m)$ time. The total length of the paths is bounded by $O(L(n,m) + I(n,m))$, and their total combinatorial complexity is bounded by $O(K(n,m) + I(n,m))$.*

*Proof.* We start with fulfilling the prerequisites of the algorithm. Finding revolving areas for all start and target positions is done in $M(n,m)$ time, and computing a free path for each robot, for all the robots, is done in $F(n,m)$ time.

For any $1 \leqslant i \leqslant m$, we compute the intersection points of $\gamma_i$ with $\partial\mathcal{C}_z$ for all $z \in Z_i$ and store these points in a sorted list $L_i^1$, sorted according to their order of appearance in $\gamma_i$. If a point is both an exit point from some $\mathcal{C}_{z_1}$ and an entrance point to a different $\mathcal{C}_{z_2}$, then we add two consecutive entries to the list, the first one for the exit point and the second one for the entrance point. In addition, for each intersection point in $L_i^1$ we keep a pointer to the last intersection point with the same $\mathcal{C}_z$. The running time of this step is $O(Q(n,m) + I(n,m)\log(I(n,m)) + m)$.

Then, we modify each path $\gamma_i$ into a new path $\gamma'_i$ that does not pass through $\mathcal{C}_z$ for all $z \in Z_i$, as described in Section 2.4. Recall that we add arcs of $\gamma_i$ until we reach an entrance point $p_z$ to some $\mathcal{C}_z$ (or reach $t_i$, the end of $\gamma_i$, and terminate). Then we add a geodesic arc on the boundary of $\mathcal{C}_z$ from $p_z$ to $q_z$, where $q_z$ is the last exit point of $\gamma_i$ from $\mathcal{C}_z$, and repeat the process recursively for the remainder of $\gamma_i$. For this purpose we use the list $L_i^1$. Note that the first point in $L_i^1$ is the first entrance point $p_z$ (or else $L_i^1$ is empty), and together with $p_z$ we keep a pointer to the respective exit point $q_z$. Moreover, the next entrance point is the successor of $q_z$ in $L_i^1$, and this invariant holds for all recursive iterations. Therefore, the running time of this step for a single path is proportional to the complexity of the given path $\gamma_i$ and the length of its intersection list $L_i^1$, thus the running time of this step for all paths is $O(K(n,m) + I(n,m))$. The complexity of each new path $\gamma'_i$ is $k(\gamma'_i) = k(\gamma_i) + O(|L_i^1|)$, since it is composed of arcs of $\gamma_i$ and at most $O(|L_i^1|)$ additional geodesic arcs.

For each path $\gamma'_i$, we compute its intersection points with $\partial\mathcal{B}_z$ for all $z \in Z_i$ and store these points in a sorted list $L_i^3$, sorted according to their order of appearance in $\gamma'_i$ from start to target. The running time of this step is $O(Q(n,m) + I(n,m)\log(I(n,m)))$.

Next, for every $1 \leqslant i \leqslant m$ we compute a reparametrization $\gamma''_i$ of $\gamma'_i$. It can be done in linear time in the complexity of the path $\gamma'_i$ and the length of the intersection list $L_i^3$ in the following manner. Let $\delta$ be such that $\frac{1}{\delta}$ equals the number of arcs in $\gamma'_i$ plus the length of $L_i^3$. We fix the time frame of each arc of $\gamma''$ to be $\delta$, and for each intersection of $\gamma'$ with $\partial\mathcal{B}_z$ for some $z \in Z_i$ we increase by $\delta$ the length of the time frame in which the image of $\gamma''_i$ is this intersection point. Namely, while moving along the path $\gamma_i''$, the robot $r_i$ "stays" at the intersection point for a time frame of length $\delta$. We treat each such time frame of length $\delta$ as an arc of $\gamma_i''$, therefore the complexity of $\gamma_i''$ is $k(\gamma_i) + O(|L_i^1|+|L_i^3|)$. For all paths, the running time of this part of the algorithm is $O(K(n,m) + I(n,m))$.

It is only left to add retraction paths. For each $1 \leqslant i \leqslant m$ we add all the retraction paths caused by $\gamma_i''$ simultaneously, in a single process. We track $\gamma_i''$ by sequentially processing each of its arcs, while maintaining a set $\mathcal{H} \subseteq Z_i$ of occupied start and target positions that interfere with $\gamma_i''$. In the beginning $\mathcal{H}$ is empty. Recall that $\gamma_i''$ has three types of arcs: (1) Entrance points to any $\mathcal{B}_z$, namely, intersection points in which the robot $r_i$ following $\gamma_i''$ is about to start intersecting $\mathcal{A}_z$. (2) Exit points from any $\mathcal{B}_z$, namely, intersection points in which the robot $r_i$ following $\gamma_i''$ has just stopped intersecting $\mathcal{A}_z$. (3) Sub-arcs of $\gamma_i''$. For an arc that is an entrance point $p$ to some $\mathcal{B}_z$, we add $z$ to $\mathcal{H}$, and start a new retraction path from $z$ by creating the segment from $z$ to $\rho_z(p)$. Similarly, for an arc that is an exit point $p$ from some $\mathcal{B}_z$, we remove $z$ from $\mathcal{H}$, and finish the retraction path from $z$ by creating the segment from $\rho_z(p)$ to $z$. For an arc $e$ that is a sub-arc of $\gamma_i'$, for each $z \in \mathcal{H}$ we extend its retraction path by creating the arc $\mathcal{P}_{e,z}$. Note that the size of $\mathcal{H}$ is always $O(1)$, since each point on the path is at distance at most $4$ from $O(1)$ centers of revolving areas (by Lemma 1). Therefore the running time is $O(1)$ for each arc of $\gamma_i''$, and the running time of this step for every $1 \leqslant i \leqslant m$ is proportional to the complexity of $\gamma_i''$, which is bounded by $O(k(\gamma_i)+|L_i^1|+|L_i^3|)$. Thus the running time for all paths together is $O(K(n,m)+I(n,m))$.

Therefore the total running time of the algorithm is

$$O(M(n,m) + F(n,m) + Q(n,m) + I(n,m)\log(I(n,m)) + K(n,m) + m).$$

As mentioned above, the complexity of $\gamma_i''$ is $k(\gamma_i) + O(|L_i^1|+|L_i^3|)$, for any $1 \leqslant i \leqslant m$. The length of $\gamma_i''$ is the length of $\gamma_i$ plus $O(|L_i^1|)$, since we add at most $O(|L_i^1|)$ geodesic arcs of radius 1. Thus the total complexity of all $\gamma_i''$ is $O(K(n,m) + I(n,m))$ and their total length is $O(L(n,m) + I(n,m))$. In order to find an upper bound on the overall length and complexity of the final paths, we should bound the length and complexity of the retraction paths. At each intersection of some $\gamma_i''$ and some $\mathcal{B}_z$, we begin or end a retraction path by a segment of $O(1)$ length. These are $O(I(n,m))$ segments whose total length is $O(I(n,m))$. Consider a maximal sub-path $\tilde{\gamma}$ of some $\gamma_i''$ inside some $\mathcal{B}_z$, and consider the retraction path $\mathcal{P}_{\tilde{\gamma},z}$. The complexity of $\mathcal{P}_{\tilde{\gamma},z}$ equals the complexity of $\tilde{\gamma}$. In addition, since $\tilde{\gamma}$ does not intersect $\mathcal{C}_z$, it follows from the definition of the retraction that the length of $\mathcal{P}_{\tilde{\gamma},z}$ is at most the length of $\tilde{\gamma}$. Summing over all the retraction paths, their total length is $O(L(n,m) + I(n,m))$ and their total complexity is $O(K(n,m) + I(n,m))$, since each arc of a path is contained in $O(1)$ balls $\mathcal{B}_z$. Therefore, the total length of the final paths $\pi_i$, $1 \leqslant i \leqslant m$, is $O(L(n,m) + I(n,m))$, and their total complexity is $O(K(n,m) + I(n,m))$.

### 4.1 Parallel Computation

We point out that our framework is "embarrassingly parallel" [36]. Initial paths and revolving areas can be computed in parallel for all robots, since they do not depend on each other. Once all initial paths and revolving areas are fixed, for each $1 \leqslant i \leqslant m$, modifying the $i$'th path and computing the retraction paths it incurs can also be done in parallel for each $i$, since it depends only on the initial paths and the revolving areas.

In Section 3 we suggested concrete methods for operations that are used in the algorithm as "black boxes". All of these methods can be executed for each robot independently (except for the preprocessing stage in the general method for finding revolving areas), and thus each method can be executed for all robots in parallel.

## 5 The Case of Shortest Paths in $\mathbb{R}^2$

We analyze the special case of shortest paths in $\mathbb{R}^2$. We show upper bounds on the parameters we use in the general analysis and achieve a more refined analysis for this case.

As mentioned in Section 3, in $\mathbb{R}^2$ we can find a shortest path for a single unit-disc robot moving among polygonal obstacles of complexity $n$ in $O(n^2)$ time, and it is well known that the complexity of such a path is $O(n)$. Thus $F(n,m) = O(n^2 m)$ and $K(n,m) = O(nm)$. By the methods suggested in Section 3, all revolving areas can be computed in $M(n,m) = O(n \log^2(n) + mn)$ time, and the intersection points between a single shortest path $\gamma$ and a set of circles $\mathcal{C} \subseteq \{\partial \mathcal{C}_z, \partial \mathcal{B}_z \mid z \in S \cup T\}$ can be computed in $O((n+m)\log(n+m) + I(\gamma, \mathcal{C}))$ time, where $I(\gamma, \mathcal{C})$ is the number of intersections. We now give upper bounds on the number of intersections and the time it takes to compute them.

Recall that for each $1 \leqslant i \leqslant m$ we compute the intersections between the shortest path $\gamma_i$ and $\partial \mathcal{C}_z$ for all $z \in Z_i$, and the intersections between the modified path $\gamma_i'$ and $\partial \mathcal{B}_z$ for all $z \in Z_i$. Each $\gamma_i$ intersects each $\partial \mathcal{C}_z$ at most twice, because revolving areas do not intersect obstacles. Each modified path $\gamma_i'$ is composed of arcs of $\gamma_i$ and additional $O(m)$ circular arcs of $\partial \mathcal{C}_z$. Each such circular arc intersects at most $O(1)$ circles $\partial \mathcal{B}_z$, by Lemma 1. In the next lemma we show that the number of intersections between $\gamma_i$ and all circles $\partial \mathcal{B}_z$, $z \in Z_i$, is $O(n+m)$. Therefore, the total number of intersections is $I(n,m) = O(nm + m^2)$, and they can be computed in $Q(n,m) = O(m(n+m)\log(n+m))$ time.

**Lemma 6.** *Let $\gamma$ be a shortest path of a unit-disc moving among polygonal obstacles, and let $Z$ be a set of $m$ centers of revolving areas. Then the number of intersections between $\gamma$ and all $\partial \mathcal{B}_z$, $z \in Z$, is $O(k(\gamma) + m)$.*

*Proof.* Let $z \in Z$. Denote by $k_z(\gamma)$ the number of breakpoints of $\gamma$ in $B_6(z)$, where $B_6(z)$ is a disc of radius 6 centered at $z$. We show that $\gamma$ intersects $\partial \mathcal{B}_z$ at most $4k_z(\gamma) + O(1)$ times. Each arc of $\gamma$ can intersect $\partial \mathcal{B}_z$ at most twice, so it is sufficient to show that no more than $2k_z(\gamma) + O(1)$ arcs intersect $\partial \mathcal{B}_z$. At most $2k_z(\gamma)$ of these arcs have an endpoint in $B_6(z)$. Consider the arcs that intersect $\partial \mathcal{B}_z$ and both of their endpoints are not in $B_6(z)$. Notice that these arcs must be line segments, and denote the set of these segments by $\mathcal{U}$. For any $u \in \mathcal{U}$ we define $\mathcal{N}_u := \{x \in \mathcal{B}_z \mid \|x - u\| < \frac{1}{2}\}$. Let $u_1, u_2 \in \mathcal{U}$, and assume towards a contradiction that $\mathcal{N}_{u_1} \cap \mathcal{N}_{u_2} \neq \emptyset$. Let $x \in \mathcal{N}_{u_1} \cap \mathcal{N}_{u_2}$. There exist $x_1 \in u_1$ and $x_2 \in u_2$ such that $\|x_1 - x\| < \frac{1}{2}$ and $\|x_2 - x\| < \frac{1}{2}$. Thus $\|x_1 - x_2\| < 1$, and in particular $\|x_1 - u_2\| < 1$. Since $x_1 \in \mathcal{F}$ and $u_2 \subseteq \mathcal{F}$, the segment from $x_1$ to $u_2$ that is perpendicular to $u_2$ is contained in $\mathcal{F}$. See Figure 5. Therefore $\gamma$ can be shortened using this segment, which is a contradiction. We conclude that $\mathcal{N}_{u_1} \cap \mathcal{N}_{u_2} = \emptyset$ for any $u_1, u_2 \in \mathcal{U}$, and thus $|\mathcal{U}| \leqslant \frac{\mathrm{vol}(\mathcal{B}_z)}{\min_{u \in \mathcal{U}} |\mathcal{N}_u|} \leqslant const.$[2] Therefore there are at most $O(1)$ such segments, and indeed the overall number of intersections between $\gamma$ and $\partial \mathcal{B}_z$ is at most $4k_z(\gamma) + O(1)$.

Each breakpoint of $\gamma$ is at distance smaller than 6 from $O(1)$ centers of revolving areas, since the distance between every two centers of revolving areas is at least 2 (Lemma 1). Thus $\sum_{z \in Z} k_z(\gamma) = O(k(\gamma))$, and the total number of intersections between $\gamma$ and $\partial\mathcal{B}_z$ summed over all $z \in Z$ is $\sum_{z \in Z} O(k_z(\gamma) + 1) = O(k(\gamma) + m)$.

[2] Moreover, it follows that for any $u_1, u_2 \in \mathcal{U}$ the distance between any of their intersection points with $\partial \mathcal{B}_z$ is at least 1, therefore $|\mathcal{U}| \leqslant \lfloor 6\pi \rfloor = 18$.

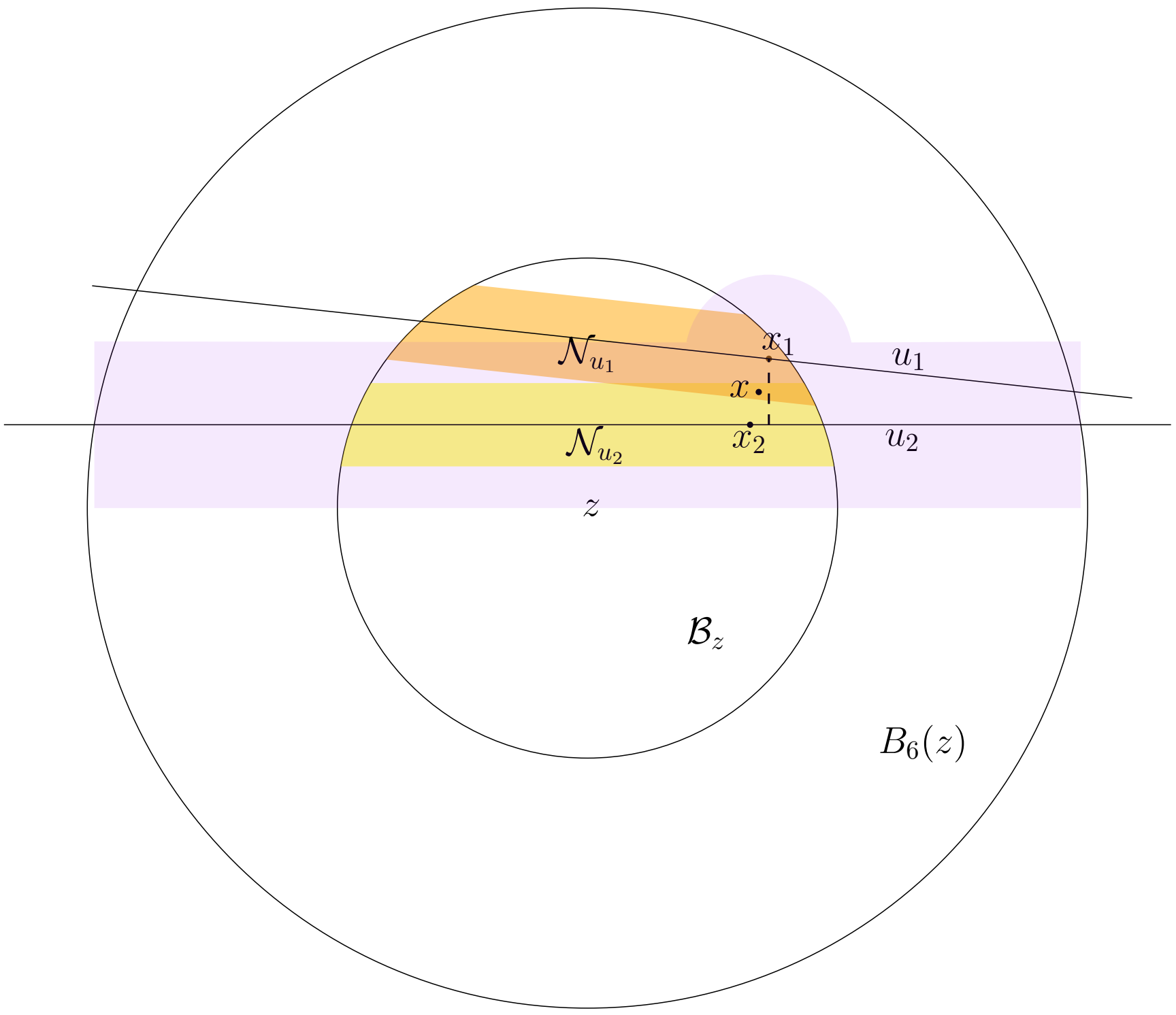

**Fig. 5:** Illustration of the proof of Lemma 6. The purple area does not contain obstacles, since $x_1 \in \mathcal{F}$ and $u_2 \subseteq \mathcal{F}$. Therefore the segment from $x_1$ to $u_2$ that is perpendicular to $u_2$ (dashed) is free.

We conclude the following theorem.

**Theorem 2.** *The algorithm can compute free paths for $m$ unit-disc robots in $\mathbb{R}^2$ moving among polygonal obstacles of complexity $n$ in $O(n^2 m + m(m+n)\log(m+n))$ time. The total length of the paths is bounded by $O(L(n,m) + nm + m^2)$, where $L(n,m)$ is the total length of the initial $m$ shortest paths, and their total combinatorial complexity is bounded by $O(nm + m^2)$.*

### 5.1 Construction of a Bad Input for Shortest Paths in $\mathbb{R}^2$

We show an example input for which, when using shortest paths as the original paths, the algorithm computes paths whose total length is $\Theta(L(n,m) + nm + m^2)$. This matches the upper bound proven in Theorem 2.

First, notice that there is an addition of $\Theta(m^2)$ to the total length when all the start and target positions are placed on a single line, in a manner that for any $i < j$ the positions $s_j$ and $t_j$ are between $s_i$ and $t_i$ (see Figure 6a).

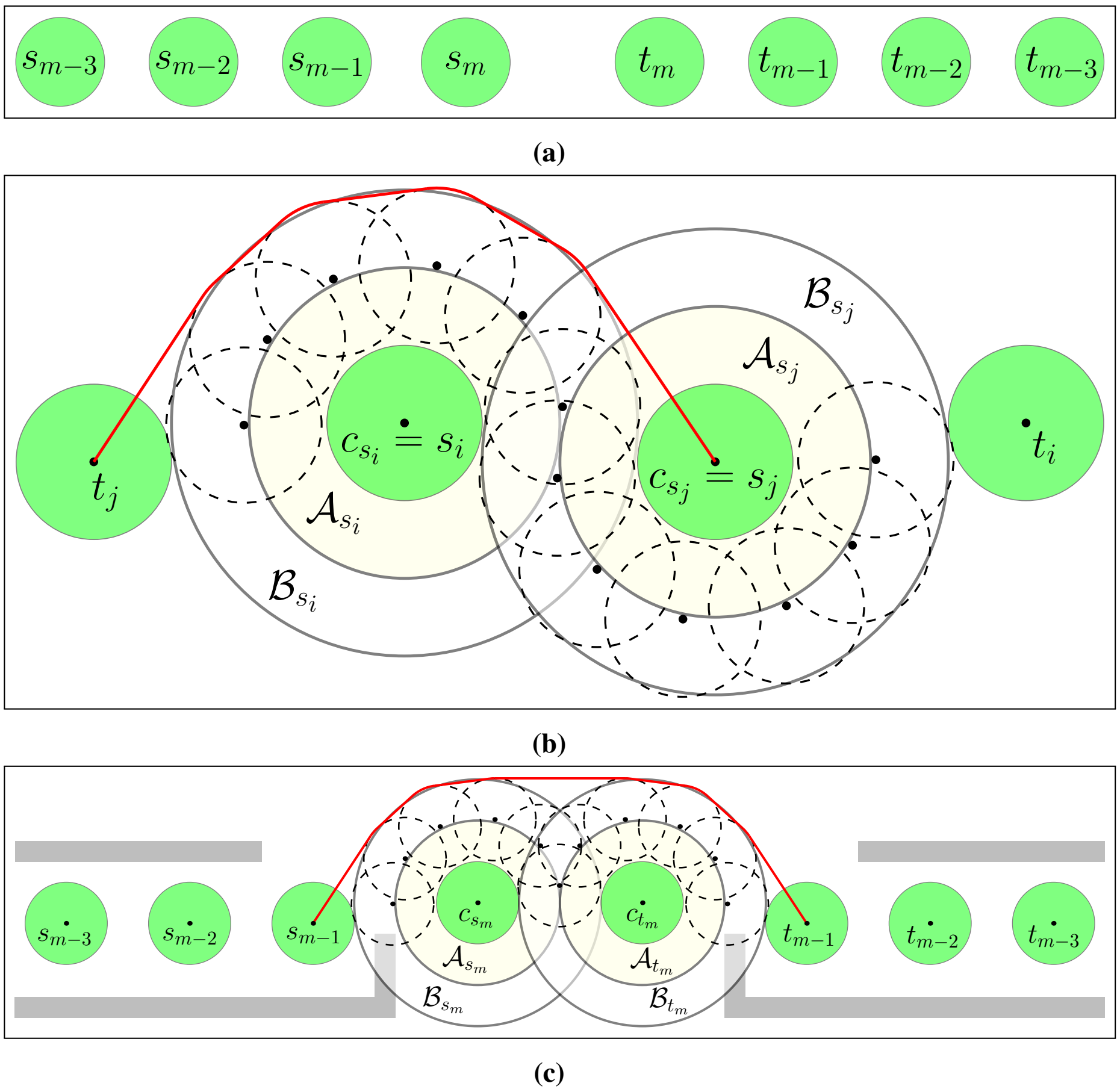

**Fig. 6:** Examples of inputs for which the total length of the paths increases by: (a) $\Theta(m^2)$; (b) $\Theta(n)$ (with only 2 robots); (c) $\Theta(m^2 + nm)$

In order to create a scenario in which the total length of the paths increases by $\Theta(nm)$, we should have $\Theta(nm)$ intersections between the paths and the circles $\partial\mathcal{B}_z$. Since we are dealing with shortest paths and obstacles cannot intersect revolving areas, we enforce these intersections by placing many small obstacles near revolving areas. In particular, we show an example in which tiny obstacles are placed near two revolving areas, in a manner that as $n$ tends to infinity, the number of those intersections tends to infinity, and the size of each obstacle tends to zero.

Assume we have two robots, $i$ and $j$, whose start and target positions are $t_j = (0, 0)$, $s_i = (4, \frac{1}{2})$, $s_j = (8, 0)$, $t_i = (12, \frac{1}{2})$. We choose each start or target position as the center of its revolving area. For simplicity we let the obstacles be points, that can be later replaced with tiny triangles. We place $\frac{n}{2}$ point obstacles on the upper (lower) half of the circle of radius $2 + \varepsilon$ centered at $c_{s_i}$ ($c_{s_j}$), for infinitesimally small $\varepsilon > 0$. We place the points in a way that the distance between any two consecutive points is smaller than 2; see Figure 6b.

Due to symmetry, we can assume without loss of generality that $j$ moves before $i$. Each point obstacle prevents the robots to move through a disc of radius 1 centered at the obstacle. Consider the circles that are the boundaries of these discs. The shortest path of $j$ passes above the revolving area $\mathcal{A}_{s_i}$, and it is composed of arcs of these circles alternating with parts of the common tangents of any two consecutive circles. Since the obstacles are placed at distance $2+\varepsilon$ from $c_{s_i}$, for small enough $\varepsilon$, this path contains $\Theta(n)$ arcs that are not in $\mathcal{B}_{s_i}$, and $\Theta(n)$ parts of the tangents that are inside $\mathcal{B}_{s_i}$. Therefore the number of intersections of the shortest path of $j$ with the boundary of $\mathcal{B}_{s_i}$ is $\Theta(n)$, and the total length of all retraction paths is $\Theta(n)$.

The two examples we presented can be combined into a scenario for which the algorithm creates paths whose total length is $\Theta(L(n,m)+nm+m^2)$. We create a modified version of the second example, such that both the start and target positions of some robot, say robot $m$, are surrounded by tiny obstacles as in that example, but all the obstacles are on the upper halves of the circles. We place this structure between the start positions and the target positions of the first example, and add a few rectangular obstacles to enforce the paths of $1,\ldots,m-1$ to pass near $\Theta(n)$ of the tiny obstacles. See Figure 6c. For any $1 \leqslant i < m$, the path of $i$ intersects both $\partial\mathcal{B}_{s_j}$ and $\partial\mathcal{B}_{t_j}$ for any $i < j < m$ (one of them must contain the robot $j$), and has $\Theta(n)$ intersections with both $\partial\mathcal{B}_{s_m}$ and $\partial\mathcal{B}_{t_m}$ (one of them must contain the robot $m$). Therefore the total length of all retraction paths is $\Theta(nm+m^2)$ as desired.

## 6 Finding Optimal Order of Paths is NP-Hard

Recall that our algorithm first finds a path for each robot, disregarding possible interferences with other robots, and then modifies these paths to avoid collisions. In particular, a path is locally modified when it passes near a start or target position that is occupied by another robot during the motion. Since each such interference has to be handled, the number of these interferences directly affects the running time, as well as the length and complexity of the modified paths. The order in which the paths are executed might have a major effect on the number of interferences, as a path might pass near a start position of some robot but not near its target position, or vice versa. For example, in Version II of the Tunnel scenario, which is presented in the experiments chapter, when executing the paths in the given order, $\Theta(m^2)$ interferences have to be handled, whereas there exists a different order of execution for which all the original paths do not incur any collision. In this scenario, the difference in the running time and the total length and complexity of the paths when using the various orders is significant. (See Section 7 for full explanation.) Therefore, a natural extension of our algorithm would be to find an order of execution of the paths in which the number of interferences is minimal. However, we show that finding such an order is NP-hard.

The hardness result that we now prove applies more generally to any algorithm for motion-coordination of many robots that first finds a path for each robot (disregarding other robots) and then handles possible interferences. Therefore we first state the problem in its general form. To simplify notation, we define the problem for unit-balls, though its extension to any type of bounded robots is straightforward.

### 6.1 Problem Definition and Hardness Proof

We assume that each start or target position $z$ has a *neighborhood*, which is a bounded subset of $\mathbb{R}^d$ that contains the unit-ball $\mathcal{D}_z$, such that a path *interferes* with $z$ if the path passes

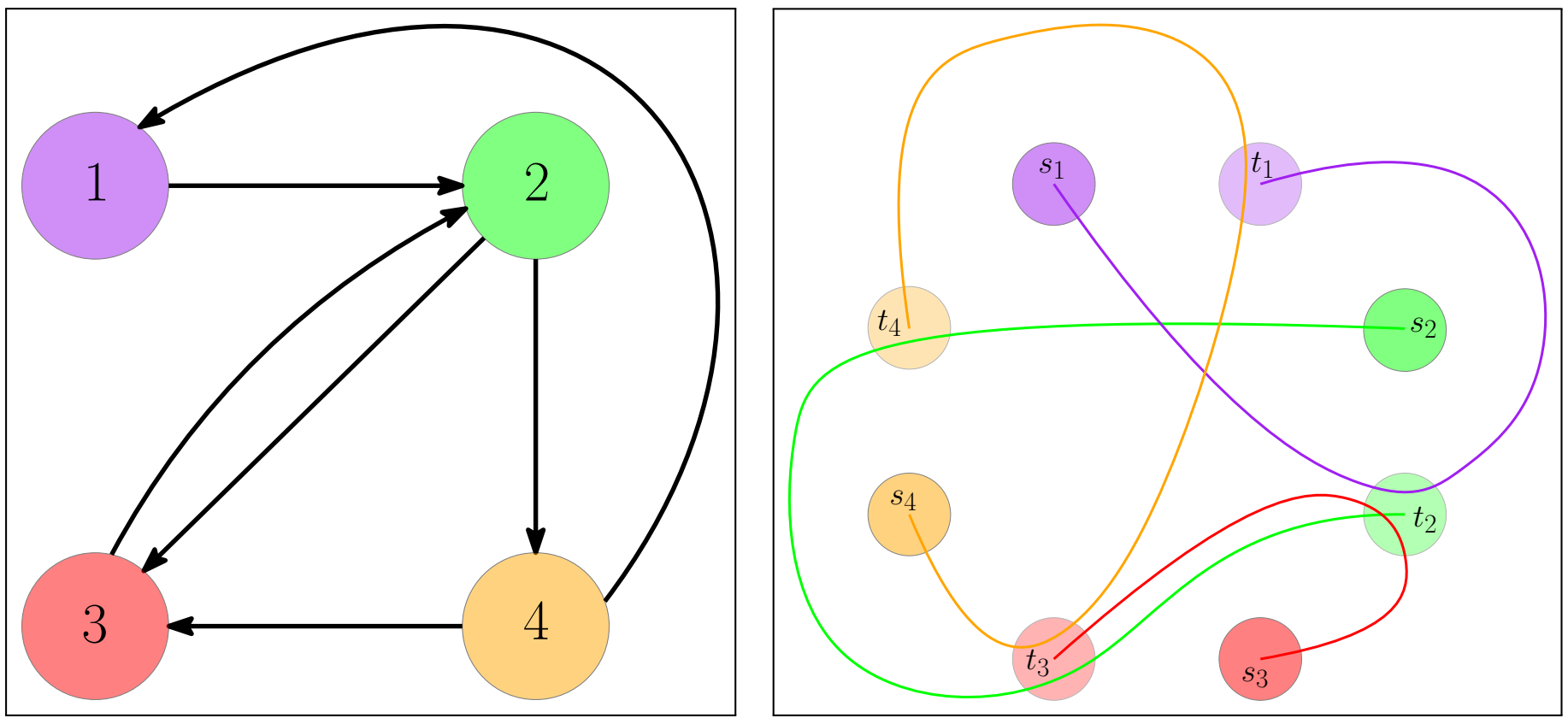

**(a)** a graph **(b)** paths created by applying the reduction through the neighborhood of $z$. We assume that each neighborhood is bounded by a ball of radius $\ell$, for some $\ell$ that does not depend on the locations of the start and target positions. The specific subset that serves as a neighborhood depends on the algorithm being used. For example, in our algorithm, the neighborhood of a start or target position $z$ is $\mathcal{B}_z$, the ball of radius 3 centered at the center of its revolving area, and $\ell = 3$.

**Fig. 7:** An example for applying the reduction from the proof of Theorem 3 on a graph. For any vertex of the graph, its path passes through the targets associated with its outgoing neighbors.

Assume that we have a set $\Gamma$ of $m$ paths, $\gamma_i$ from $s_i$ to $t_i$, for $1 \leqslant i \leqslant m$. Each path might interfere with any start or target position (perhaps several times), and any such interference has to be locally resolved. We wish to find a permutation of the paths, such that if the paths are executed one after the other according to this order, the total number of interferences is minimal.

We define the *interference graph* of $\Gamma$ as follows. Let $G_\Gamma$ be a directed graph on $m$ vertices. For any $i \neq j$, for each interference of $\gamma_i$ with $t_j$, and for each interference of $\gamma_j$ with $s_i$, we add an edge $(i, j)$ to $G_\Gamma$. Note that since any path might interfere with any start or target position multiple times, the resulting graph may be a multi-graph.

Let $\sigma : \{1, \ldots, m\} \to \{1, \ldots, m\}$ be a permutation. Then the number of interferences that have to be handled when the order of execution is $\gamma_{\sigma(1)}, \ldots, \gamma_{\sigma(m)}$ equals the number of edges $(\sigma(i), \sigma(j))$ such that $\sigma(i) > \sigma(j)$.

Therefore finding an order of execution that minimizes the number of interferences is equivalent to finding a *minimum feedback arc set* [37], namely finding a permutation of the nodes in which the number of edges that have to be removed in order to get a topological sort is minimal. The "minimum feedback arc set" problem is known to be NP-hard [16].

**Theorem 3.** *Finding an order of execution of the paths in which the number of interferences is minimal is NP-hard.*

*Proof.* We show that for any simple, directed, unweighted graph $G(V, E)$, there exists a set of paths $\Gamma$ such that the interference graph of $\Gamma$ is $G$. Given $G$, we define the paths $\Gamma$ in the following way. Denote the vertices of $G$ by $V = \{1, \ldots, m\}$. For any $v \in V$ let $s_v, t_v$ be two points in $\mathbb{R}^d$, such that the distance between any two of the $2m$ points $\{s_v, t_v\}_{v \in V}$ is larger than $2\ell$. For any $v \in V$, let $I_v = \{u \in V \mid (v, u) \in E\}$ be the set of its outgoing neighbors. For any $v \in V$, let $\gamma_v$ be a path from $s_v$ to $t_v$, such that $\gamma_v$ passes through the unit-ball centered at $t_u$ for any $u \in I_v$, and does not pass through the ball of radius $\ell$ centered at any $u \notin I_v$. In particular, $\gamma_v$ interferes with the target positions of the outgoing neighbors of $v$, and does not interfere with any other start or target position. See Figure 7. Then $G_\Gamma = G$, which proves our claim.

### 6.2 Heuristic for Determining the Order

In the experiments (Chapter 7), we test the following heuristic for determining the order of the paths. We construct two interference graphs, for two possible meanings of "interference". For the first graph, we say that a revolving area $\mathcal{A}_z$ interferes with a path if the path passes through $\mathcal{B}_z$, and for the second graph, if the path passes through $\mathcal{C}_z$. Notice that any strongly connected component (SCC) of the second graph, is contained in an SCC of the first graph. For each graph, we find a topological order of its SCCs. We define the following order of the paths. Let $\sigma$ be a random permutation of $1, \ldots, m$. We say that $i$ is before $j$ if one of the following holds: (1) the SCC of $i$ appears before the SCC of $j$ in the first graph; (2) $i$ and $j$ belong to the same SCC in the first graph and the SCC of $i$ appears before the SCC of $j$ in the second graph; (3) $i$ and $j$ belong to the same SCCs in both graphs and $i$ appears before $j$ in $\sigma$.

We expect this heuristic to be helpful in scenarios in which the interference graph is "close" to having a topological order, namely, when the number of SCCs is relatively large. Indeed, it can be seen in the experimental results in Section 7 that the heuristic considerably improves the results of the Tunnel scenario (Version II) mentioned above.

## 7 Experiments

We implemented our algorithm for the case of unit-disc robots in the plane moving amidst polygonal obstacles and tested it on several scenarios.

### 7.1 Implementation Details

The algorithm was implemented in C++ using CGAL [34] and Boost [30]. The code was tested on a PC with Intel i7-6700HQ 2.60GHz processor with 12GB of memory, running a Windows 10 64-bit OS.

We chose to implement the operations from Section 3 as follows. For finding original paths, we find for each robot its shortest path (disregarding other robots) using a visibility graph. We use the general method described in Section 3.1 to find revolving areas.[3]

For each robot, we compute the intersection points of its path with the circles $\partial\mathcal{C}_z$ and $\partial\mathcal{B}_z$ for any center of an occupied revolving area $z$, and construct the modified path in the following manner. We compute an arrangement (using CGAL Arrangements [10]) containing all the arcs of the path and the relevant circles. Note that all the required intersections are vertices of this arrangement, so by following the arcs of the path we can get these intersection points sorted according to their order along the path. We construct the final paths in two steps. First, we traverse the arcs of the path, while marking the edges of the arrangement through which we pass, until we reach the end of the path or any intersection with some $\partial\mathcal{C}_z$. In the latter case, we follow the boundary of $\mathcal{C}_z$, and mark the edges composing it in the arrangement, until we reach its second intersection with the path. Then we continue traversing the path. When we reach the end of the path, the marked edges compose the modified path, as described in Section 2.4. Afterwards, we follow the modified path from start to target, while keeping track of the discs $\mathcal{B}_z$ that contain each edge or vertex, and constructing the retraction paths as described in Section 4.

[3] We emphasize that this method is used for all scenarios, even where it is clear that each start or target position can serve as a center for its revolving area.

The heuristic for choosing the order of the paths, which is described in Section 6.2, is implemented in a similar way, by constructing for each path an arrangement containing the edges of the path and all the circles $\partial\mathcal{B}_z$ and $\partial\mathcal{C}_z$ for finding the intersections.

Excluding the heuristic for choosing the order of the paths, the implementation is deterministic and parameter free.

### 7.2 Experiments

We tested our code on the scenarios described below. We tested each scenario with and without using the heuristic for choosing the order of the paths. We report the running times for various numbers of robots. We also report the ratio between the total length of the paths and the total length of the original shortest paths, where the latter is a (usually not tight) lower bound on the total length of the paths in an optimal solution. We call this ratio *dist ratio* in the tables below. The results appear in Tables 1 – 4.

*Grid:* (See Figure 8a.) The start and target positions are organized in a grid in a square room, bounded by walls, where the start positions are in the upper half of the square and the target positions are in the lower half. The start and target positions are assigned to the robots by a random permutation. The distance between any two adjacent start or target positions is $3$, which is the minimal distance required for existence of revolving areas.

*Triangles:* (See Figure 8b.) All obstacles are triangles, which, as well as the start and target positions, are chosen at random.

*Tunnel - Version I:* (See Figure 8c.) In this scenario the robots have to move from the upper part of a narrow winding tunnel to its lower part. For any $i < j$, the start position of $i$ is closer than the start position of $j$ to the upper end of the tunnel (top-left corner), and the target of $i$ is closer than the target of $j$ to the lower end of the tunnel (bottom-left corner). Therefore $i$ has to bypass $j$, and the tunnel is designed in a manner that for any path of $i$ from start to target, it is impossible to move $j$ to a position in which it does not block the path of $i$. Therefore $\binom{m}{2}$ interactions between the robots are required, so the total length of the paths has to be greater than the total length of all original paths by $\Omega(m^2)$.

*Tunnel - Version II:* (See Figure 8d.) This scenario is the same as Version I, but the order of the target positions is reversed. Namely, for any $i < j$, the start position of robot $i$ is closer than the start position of robot $j$ to the upper end of the tunnel, but the target of $i$ is farther than the target of $j$ from the lower end. Therefore, if the algorithm computes paths using the given order of the robots, then $\binom{m}{2}$ interactions are required, just as in Version I. However, there exists a different order, the reverse order, in which the original paths are free paths, so if the algorithm chooses this order, it finds the optimal solution.

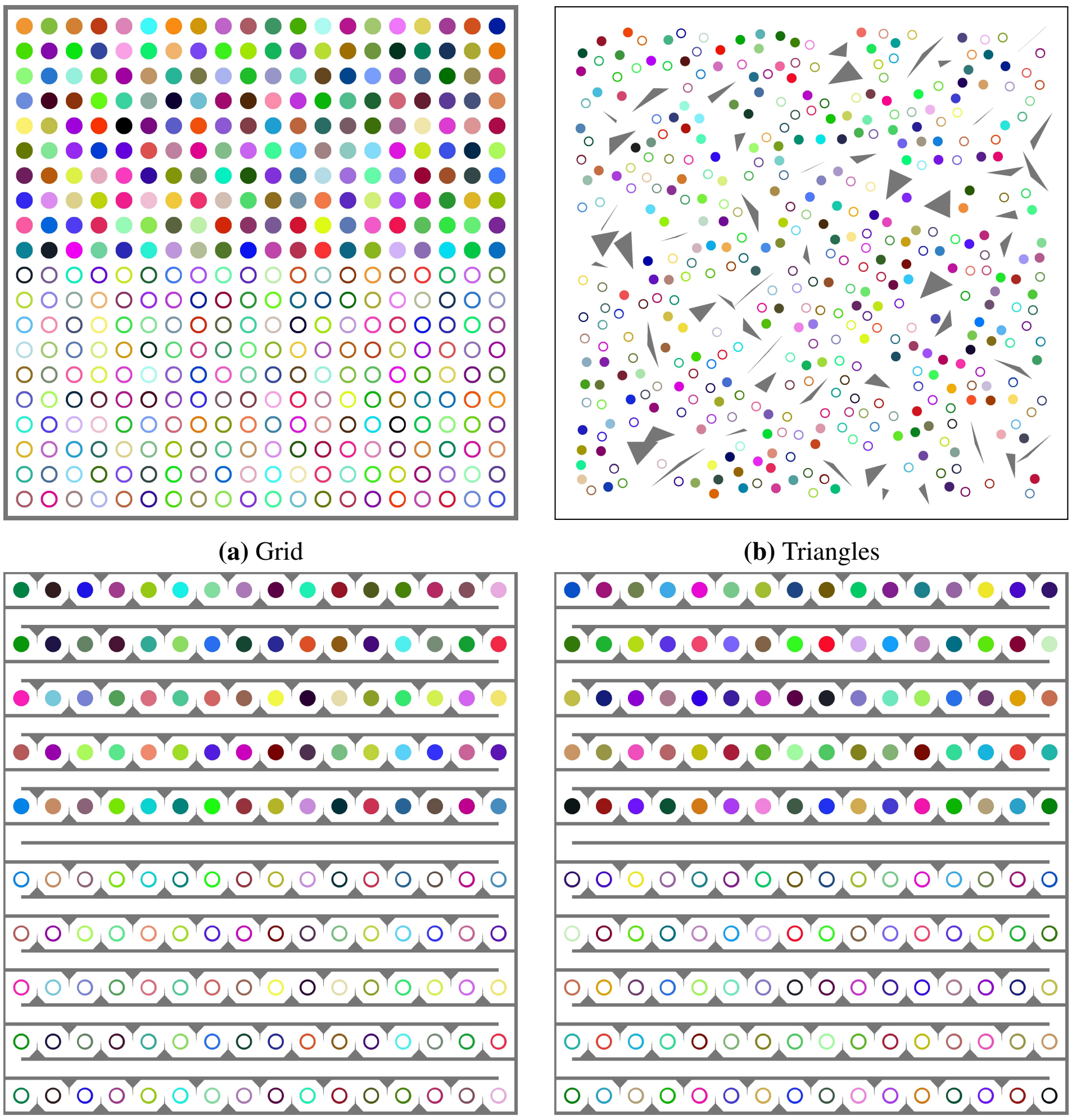

(a) Grid

(b) Triangles

(c) Tunnel - Version I

(d) Tunnel - Version II

**Fig. 8**

### 7.3 Conclusions

Our experiments show that our algorithm is efficient in practice. All the scenarios that we tested are very dense and of high combinatorial complexity, and we managed to solve them for hundreds of robots in a short amount of time.

The Grid and Triangles scenarios are modified versions of scenarios that were tested by Solovey et al. [32]. In their work, they solved the unlabeled version under separation assumptions that are stronger than our assumptions (that is, any input that satisfies their assumptions also satisfies ours). In [32] they reported that they solved the Grid scenario for $40$ robots in $311$ seconds, and the Triangles scenario for $32$ robots and $38$ triangles in $6394$ seconds. In our experiments, we decreased the distances between the start and target positions and the obstacles to match the minimal requirements of our algorithm, making the scenes as dense as possible, and assigned random labels to the start and target positions.[4] For a similar number of robots and triangles as in [32], we solve both the Grid and Triangles scenarios significantly faster, in only a few seconds for each.

In all the scenarios that we tested, the ratio between the total length of the final paths and the total length of the original shortest paths is a small number, specifically smaller than 3. We showed in Section 4 that the total length of the final paths is $O(L(n, m) + I(n, m))$, where $L(n, m)$ is the total length of the original paths and $I(n, m)$ is the total number of intersections between the paths and all the circles $\partial\mathcal{B}_z$ and $\partial\mathcal{C}_z$. The results of the experiments suggest that the constant hidden in the big-$O$ notation is small, and that the contribution of $O(I(m, n))$ to the total length is not significant in these scenarios.

In Section 5.1 we showed an example with two robots, in which the total length of the original paths is $O(1)$ and the total length of the final paths is $\Theta(n)$, thus the distance ratio is $\Theta(n)$, and tends to infinity as $n$ tends to infinity. However, in order to construct this example we had to place many tiny obstacles, whose size tends to zero, and we do not expect this to happen in practical scenarios.

As expected, the heuristic we tested for determining the order of the paths has the most effect in Version II of the Tunnel scenario. It finds the optimal order, in which all the original shortest paths are free, and thus the algorithm finds the optimal solution, reducing the distance ratio from 1.5 to 1. In addition, since there are no interactions between the robots, less computations are required and the running time slightly decreases. In contrast, as explained earlier, in Version I of the Tunnel scenario any order of the paths requires $\binom{m}{2}$ interactions between the robots, thus the distance ratio remains unchanged. In the other scenarios, the distance ratio improves, but not significantly.

## 8 Conclusions and Further Research

As we have seen above, the implementation of our algorithm runs very fast for the two-dimensional case. As we pointed out in Section 4.1, our framework is “embarrassingly parallel” [36], which may be useful for higher-dimensional instances, since the various geometric procedures are expected to be more costly.

Any algorithm for the labeled multi-robot problem can also be used for solving the unlabeled multi-robot problem, by assigning labels to the start and target positions. As explained in Section 7.3, our test scenarios Grid and Triangles were created that way, by assigning random labels to the unlabeled scenarios. We solved these scenarios significantly faster than an algorithm by Solovey et al. [32], which was specifically designed for the unlabeled problem; notice though that they provide certain optimality guarantees on the total length of the paths that their algorithm produces. By choosing an assignment of labels in a more sophisticated manner, for example using the Hungarian algorithm [18], one could try to minimize the total length of the paths while using our algorithm as well.

[4] Note that our algorithm can always be applied to unlabeled scenarios by assigning random labels. We discuss a possible modification of our algorithm toward the unlabeled case in Section 8.

It would be interesting to extend the algorithm for other shapes of robots, such as simple polygons, and find the minimal radius of revolving areas for these robots. In particular, one can consider the case of ball robots with different radii. We can extend the notion of revolving area as follows. For any $1 \leqslant j \leqslant m$ let $\delta_j$ be the radius of robot $r_j$, and for a start or target position $z$ of robot $r_j$ let $\mathcal{D}_z$ be the ball of radius $\delta_j$ centered at $z$. Then, for any start or target position $z$ of robot $r_j$, we define a revolving area $\mathcal{A}_z$ as an open ball of radius $\ell_j := \delta_j + \max_{i \neq j} \delta_i$ that contains $\mathcal{D}_z$, does not intersect obstacles, and does not intersect $\mathcal{D}_y$ for any start or target position $y \neq z$. Note that the definitions of retraction point and retraction path, as defined in Section 2.3, remain well-defined in these settings. When robot $r_i$ follows a path $\gamma$ that does not pass through the ball of radius $\ell_j - \delta_j$ centered at $c_z$, which is the center of $\mathcal{A}_z$, then $r_j$ can follow the retraction path from $z$ with respect to $\gamma$, and this motion does not incur collision between $r_j$ and $r_i$. It remains to generalize the lemmas and their proofs for these settings. (It might be necessary to require stronger assumptions.)

In Theorem 3 we showed that finding the optimal order of the paths is NP-hard for any decoupled algorithm that locally resolves interferences. A natural question is whether it is still NP-hard for shortest paths, or for other types of realistic paths.

| Grid | | | | |
|---|---|---|---|---|
| | without heuristic | | with heuristic | |
| m | time (sec) | dist ratio | time (sec) | dist ratio |
| 20 | 1 | 1.809 | 1 | 1.414 |
| 30 | 1 | 2.047 | 1 | 1.613 |
| 50 | 3 | 2.491 | 3 | 1.804 |
| 100 | 10 | 2.385 | 15 | 2.321 |
| 200 | 37 | 2.530 | 45 | 2.564 |
| 300 | 75 | 2.546 | 98 | 2.469 |
| 400 | 134 | 2.453 | 175 | 2.425 |
| 500 | 203 | 2.591 | 265 | 2.610 |
| 1000 | 830 | 2.589 | 1081 | 2.620 |

**Table 1:** Experimental results for the Grid scenario.

| Triangles (in 100x100 square) | | | | | | |
|---|---|---|---|---|---|---|
| | 10 triangles | | 30 triangles | | 50 triangles | |
| m | time (sec) | dist ratio | time (sec) | dist ratio | time (sec) | dist ratio |
| 20 | 3 | 1.009 | 23 | 1.019 | 61 | 1.005 |
| 30 | 4 | 1.016 | 24 | 1.010 | 66 | 1.018 |
| 50 | 4 | 1.028 | 27 | 1.048 | 73 | 1.039 |
| 100 | 8 | 1.061 | 33 | 1.070 | 89 | 1.070 |
| 200 | 18 | 1.142 | 53 | 1.135 | 110 | 1.122 |
| 300 | 29 | 1.216 | 76 | 1.194 | 144 | 1.201 |
| 400 | 49 | 1.276 | 104 | 1.280 | 185 | 1.279 |
| 500 | 78 | 1.360 | 144 | 1.365 | 228 | 1.348 |
| 600 | 115 | 1.430 | 190 | 1.451 | 293 | 1.426 |
| 700 | 166 | 1.505 | 248 | 1.525 | 357 | 1.503 |
| 800 | 227 | 1.566 | 314 | 1.590 | 445 | 1.595 |
| 900 | 308 | 1.673 | 426 | 1.674 | 529 | 1.664 |
| 1000 | 392 | 1.755 | 505 | 1.761 | 652 | 1.739 |

**Table 2:** Experimental results for the Triangles scenario (without the heuristic). The results with the heuristic are omitted, as the improvement is not significant for this scenario.

| Tunnel - Version I | | | | | |
|---|---|---|---|---|---|
| | | without heuristic | | with heuristic | |
| m | n | time (sec) | dist ratio | time (sec) | dist ratio |
| 20 | 242 | 36 | 1.243 | 36 | 1.243 |
| 30 | 342 | 74 | 1.248 | 75 | 1.248 |
| 50 | 542 | 197 | 1.252 | 200 | 1.252 |
| 100 | 1042 | 783 | 1.255 | 788 | 1.255 |
| 200 | 2042 | 3324 | 1.257 | 3377 | 1.257 |

**Table 3:** Experimental results for the Tunnel (Version I) scenario.

| Tunnel - Version II | | | | | |
|---|---|---|---|---|---|
| | | without heuristic | | with heuristic | |
| m | n | time (sec) | dist ratio | time (sec) | dist ratio |
| 20 | 242 | 37 | 1.487 | 36 | 1 |
| 30 | 342 | 75 | 1.497 | 73 | 1 |
| 50 | 542 | 203 | 1.505 | 195 | 1 |
| 100 | 1042 | 1065 | 1.511 | 1019 | 1 |
| 200 | 2042 | 3373 | 1.514 | 3245 | 1 |

**Table 4:** Experimental results for the Tunnel (Version II) scenario.

## A Proof of Proposition 1

**Proposition 1.** *Let $p \in \mathcal{W}$ such that $\|x - c_{s_j}\| = 3$ and $1 \leqslant \|x - c_{s_k}\| \leqslant 3$. Then for any point $u$ in the segment $[\rho_{s_j}(p), s_j]$, it holds that $\|\rho_{s_k}(p) - u\| \geqslant 2$.*

*Proof.* For a point $a$, denote by $x_i(a)$ its $i$'th coordinate, for any $1 \leqslant i \leqslant d$. For two points $a$, $b$, we say that $a$ is to the left (right) of $b$ if $x_1(a) < x_1(b)$ ($x_1(a) > x_1(b)$). For a point $a$ that is contained in the $x_1x_2$ plane, we say that $a$ is above (below) the $x_1$-axis if $x_2(a) > 0$ ($x_2(a) < 0$).

Without loss of generality, assume that $c_{s_j}$ is in the origin, $c_{s_k}$ is placed on the positive part of the $x_1$-axis, and $p$ is contained in the $x_1x_2$ plane, on or above the $x_1$-axis. We divide the proof into cases. See Figure 9 for illustration.

(i) *Assume $\mathcal{A}_{s_j} \cap \mathcal{A}_{s_k} = \emptyset$.* For any $u \in [\rho_{s_j}(p), s_j]$, since $\mathcal{D}_{\rho_{s_k}(p)} \subseteq \mathcal{A}_{s_k}$ and $\mathcal{D}_u \subseteq \mathcal{A}_{s_j}$, it holds that $\mathcal{D}_{\rho_{s_k}(p)} \cap \mathcal{D}_u = \emptyset$.
For the rest of the cases we assume that $\mathcal{A}_{s_j} \cap \mathcal{A}_{s_k} \neq \emptyset$.

(ii) *Assume $p$ lies on the $x_1$-axis.* Since the points $p$, $c_{s_j}$ and $c_{s_k}$ lie on the same line, the assumptions $\|p - c_{s_j}\| = 3$, $1 \leqslant \|p - c_{s_k}\| \leqslant 3$ and $2 \leqslant \|c_{s_j} - c_{s_k}\| < 4$ imply that $\|c_{s_j} - c_{s_k}\| = 2$. The points $\rho_{s_j}(p)$ and $\rho_{s_k}(p)$ also lie on the $x_1$-axis, at distance 1 from $c_{s_j}$ and $c_{s_k}$, to their left, respectively. Thus $\|\rho_{s_j}(p) - \rho_{s_k}(p)\| = \|c_{s_j} - c_{s_k}\| = 2$. In addition, recalling that $\mathcal{D}_{s_j} \subseteq \mathcal{A}_{s_j} \setminus \mathcal{A}_{s_k}$ we conclude that $s_j = \rho_{s_j}(p)$, which means that the segment is a single point whose distance from $\rho_{s_k}(p)$ is exactly 2.

(iii) *Assume $p$ is above the $x_1$-axis and $x_1(p) < x_1(c_{s_j})$.* Consider the triangle $\triangle c_{s_j} p c_{s_k}$. The angle near $c_{s_j}$ is obtuse. Therefore by Pythagoras Theorem

$$\|p - c_{s_k}\|^2 \geqslant \|p - c_{s_j}\|^2 + \|c_{s_j} - c_{s_k}\|^2 \geqslant 3^2 + 2^2 > 3^2.$$

Meaning $\|p - c_{s_k}\| > 3$, which is a contradiction. Namely, such a configuration is not possible.

(iv) *Assume $p$ is above the $x_1$-axis and $x_1(c_{s_j}) \leqslant x_1(p) \leqslant x_1(c_{s_k})$.* Consider the following three hyperplanes that are perpendicular to $x_1$: $h_j$ through $c_{s_j}$, $h_k$ through $c_{s_k}$ and $h_A$ that is tangent to $\mathcal{A}_{s_k}$ through the leftmost point of its closure. Notice that $h_A$ is placed between $h_j$ and $h_k$, and that the distance between $h_k$ and $h_A$ is exactly 2. By our assumptions, $p$ is placed to the left of $h_k$, and therefore $\rho_{s_k}(p)$ is placed to the right of $h_k$. Similarly, $p$ is placed to the right of $h_j$, and therefore $\rho_{s_j}(p)$ is placed to the left of $h_j$, and in particular to the left of $h_A$.
We show that $s_j$ is also placed to the left of $h_A$. Let $c_{s_k} = (2 + \varepsilon, 0, \ldots, 0)$, where $0 \leqslant \varepsilon < 2$. Using this notation, the hyperplane $h_A$ is defined by the equation $x_1 = \varepsilon$. Note that $s_j$ belongs to the closure of $\mathcal{C}_{s_j} \setminus \mathcal{B}_{s_k}$, because $\mathcal{D}_{s_j} \subseteq \mathcal{A}_{s_j}$ and $D_{s_j} \cap \mathcal{A}_{s_k} = \emptyset$. Hence the following inequalities hold.

$$\begin{cases} (x_1(s_j) - (2+\varepsilon))^2 + \sum_{i=2}^d x_i(s_j)^2 \geqslant 9 \\ \sum_{i=1}^d x_i(s_j)^2 \leqslant 1 \end{cases}$$

$$(x_1(s_j) - (2+\varepsilon))^2 + (1 - x_1(s_j)^2) \geqslant 9$$

$$x_1(s_j)^2 - 2x_1(s_j)(2+\varepsilon) + (2+\varepsilon)^2 + 1 - x_1(s_j)^2 \geqslant 9$$

$$2x_1(s_j)(2+\varepsilon) \leqslant (2+\varepsilon)^2 - 8$$

$$x_1(s_j) \leqslant \frac{4\varepsilon + \varepsilon^2 - 4}{2(2+\varepsilon)} < \frac{4\varepsilon + 2\varepsilon^2}{2(2+\varepsilon)} = \frac{2\varepsilon(2+\varepsilon)}{2(2+\varepsilon)} = \varepsilon$$

Therefore the whole segment $[\rho_{s_j}(p), s_j]$ is placed to the left of $h_A$, and its distance from $\rho_{s_k}(p)$ is at least 2.

(v) *Assume $p$ is above the $x_1$-axis, $x_1(c_{s_k}) < x_1(p)$ and $x_1(\rho_{s_j}(p)) \leqslant x_1(s_j)$.* Denote by $R$ the closure of $\mathcal{C}_{s_j} \setminus \mathcal{B}_{s_k}$, and denote by $CH(R)$ the convex hull of $R$. Let $\tilde{x_1}$ be the $x_1$ coordinate of the rightmost points of $CH(R)$. We show the following two properties, which prove our claim: (1) $[\rho_{s_j}(p), s_j] \subseteq CH(R)$; (2) For any $u \in CH(R)$, $\|\rho_{s_k}(p) - u\| \geqslant 2$.

(1) By convexity, it is sufficient to prove that the endpoints $s_j$ and $\rho_{s_j}(p)$ belong to $CH(R)$. Indeed, $s_j \in R$, because $\mathcal{D}_{s_j} \subseteq \mathcal{A}_{s_j} \setminus \mathcal{A}_{s_k}$. Also, since $\rho_{s_j}(p)$ belongs to the boundary of $\mathcal{C}_{s_j}$, and $x_1(\rho_{s_j}(p)) \leqslant x_1(s_j) \leqslant \tilde{x_1}$, it follows that $\rho_{s_j}(p) \in CH(R)$, as $CH(R)$ is the closure of $\mathcal{C}_{s_j} \cap \{x \mid x_1 \leqslant \tilde{x_1}\}$.

(2) For any $u \in CH(R)$, we have

$$\|\rho_{s_k}(p) - u\| \geqslant |x_1(\rho_{s_k}(p)) - x_1(u)| \geqslant |x_1(\rho_{s_k}(p)) - \tilde{x_1}|,$$

so it is sufficient to show that $|x_1(\rho_{s_k}(p)) - \tilde{x_1}| \geqslant 2$.

Recall that $c_{s_j} = (0, 0, \ldots, 0)$. As before, let $c_{s_k} = (2+\varepsilon, 0, \ldots, 0)$ where $0 \leqslant \varepsilon < 1$. (If $\varepsilon \geqslant 1$, then $x_1(p) > x_1(c_{s_k}) \geqslant 3$, which contradicts the assumption $\|p - c_{s_j}\| = 3$.) The rightmost points in $CH(R)$ are the intersection points of $\partial\mathcal{B}_{s_k}$ and $\partial\mathcal{C}_{s_j}$. We compute $\tilde{x_1}$:

$$\begin{cases} (\tilde{x_1} - (2+\varepsilon))^2 + \sum_{i=2}^{d} x_i^2 = 9 \\ \tilde{x_1}^2 + \sum_{i=2}^{d} x_i^2 = 1 \end{cases}$$

$$(\tilde{x_1} - (2+\varepsilon))^2 + (1 - \tilde{x_1}^2) = 9$$

$$\tilde{x_1}^2 - 2\tilde{x_1}(2+\varepsilon) + (2+\varepsilon)^2 + 1 - \tilde{x_1}^2 = 9$$

$$2\tilde{x_1}(2+\varepsilon) = (2+\varepsilon)^2 - 8$$

$$\tilde{x_1} = \frac{2+\varepsilon}{2} - \frac{4}{2+\varepsilon}$$

Next, we find a lower bound for $x_1(\rho_{s_k}(p))$. For that purpose we start by finding an upper bound on $x_1(p)$. By our assumptions, $\|p - c_{s_j}\| = 3$ and $1 \leqslant \|p - c_{s_k}\| \leqslant 3$.

$$\begin{cases} (x_1(p) - (2+\varepsilon))^2 + \sum_{i=2}^{d} x_i(p)^2 \geqslant 1 \\ \sum_{i=1}^{d} x_i(p)^2 = 9 \end{cases}$$

$$(x_1(p) - (2+\varepsilon))^2 + 9 - x_1(p)^2 \geqslant 1$$

$$2x_1(p)(2+\varepsilon) \leqslant (2+\varepsilon)^2 + 8$$

$$x_1(p) \leqslant \frac{2+\varepsilon}{2} + \frac{4}{2+\varepsilon}$$

Since the points $p$, $c_{s_k}$ and $\rho_{s_k}(p)$ lie on the same line, and since $\|c_{s_k} - \rho_{s_k}(p)\| = 1 \leqslant \|p - c_{s_k}\|$, it follows that $|x_1(c_{s_k}) - x_1(\rho_{s_k}(p))| \leqslant |x_1(p) - x_1(c_{s_k})|$. Substituting the values we get:

$$(2+\varepsilon) - x_1(\rho_{s_k}(p)) \leqslant \left(\frac{2+\varepsilon}{2} + \frac{4}{2+\varepsilon}\right) - (2+\varepsilon)$$

$$x_1(\rho_{s_k}(p)) \geqslant (2+\varepsilon) - \frac{2+\varepsilon}{2} - \frac{4}{2+\varepsilon} + (2+\varepsilon) = \frac{3}{2}(2+\varepsilon) - \frac{4}{2+\varepsilon}$$

Therefore:

$$|x_1(\rho_{s_k}(p)) - \tilde{x_1}| \geqslant \left(\frac{3}{2}(2+\varepsilon) - \frac{4}{2+\varepsilon}\right) - \left(\frac{2+\varepsilon}{2} - \frac{4}{2+\varepsilon}\right) = 2+\varepsilon \geqslant 2$$

as required.

(vi) *Assume $p$ is above the $x_1$-axis, $x_1(c_{s_k}) < x_1(p)$ and $x_1(s_j) < x_1(\rho_{s_j}(p))$.* Denote by $l_j$ the line through $p$, $c_{s_j}$ and $\rho_{s_j}(p)$, and denote by $l_k$ the line through $p$, $c_{s_k}$ and $\rho_{s_k}(p)$. Since $x_2(c_{s_j}) = x_2(c_{s_k}) < x_2(p)$ and $x_1(c_{s_j}) < x_1(c_{s_k}) < x_1(p)$, the slopes of $l_j$ and $l_k$ are positive and the slope of $l_k$ is greater than the slope of $l_j$. Therefore $x_2(\rho_{s_j}(p)) > x_2(\rho_{s_k}(p))$. Moreover, $\rho_{s_j}(p)$ is placed in the lower left quarter of the boundary of $\mathcal{C}_{s_j}$, so for any point $a$ in the closure of $\mathcal{C}_{s_j}$ such that $x_1(a) < x_1(\rho_{s_j}(p))$, it holds that $x_2(a) > x_2(\rho_{s_j}(p))$. In particular, we assume that $x_1(s_j) < x_1(\rho_{s_j}(p))$, so $x_2(s_j) > x_2(\rho_{s_j}(p))$. Therefore, since $[\rho_{s_j}(p), s_j]$ is a straight line segment, for any point $u$ in this segment, is holds that $x_1(s_j) \leqslant x_1(u) \leqslant x_1(\rho_{s_j}(p)) \leqslant x_1(\rho_{s_k}(p))$ and $x_2(s_j) \geqslant x_2(u) \geqslant x_2(\rho_{s_j}(p)) \geqslant x_2(\rho_{s_k}(p))$. In addition, for any $3 \leqslant i \leqslant d$, our assumptions imply $x_i(\rho_{s_j}(p)) = x_i(\rho_{s_k}(p)) = 0$. Putting it all together, it holds that $|x_i(\rho_{s_k}(p)) - x_i(u)| \geqslant |x_i(\rho_{s_k}(p)) - x_i(\rho_{s_j}(p))|$ for any $1 \leqslant i \leqslant d$. Therefore, by Lemma 3, $\|\rho_{s_k}(p) - u\| \geqslant \|\rho_{s_k}(p) - \rho_{s_j}(p)\| \geqslant 2$ for any $u \in [\rho_{s_j}(p), s_j]$.

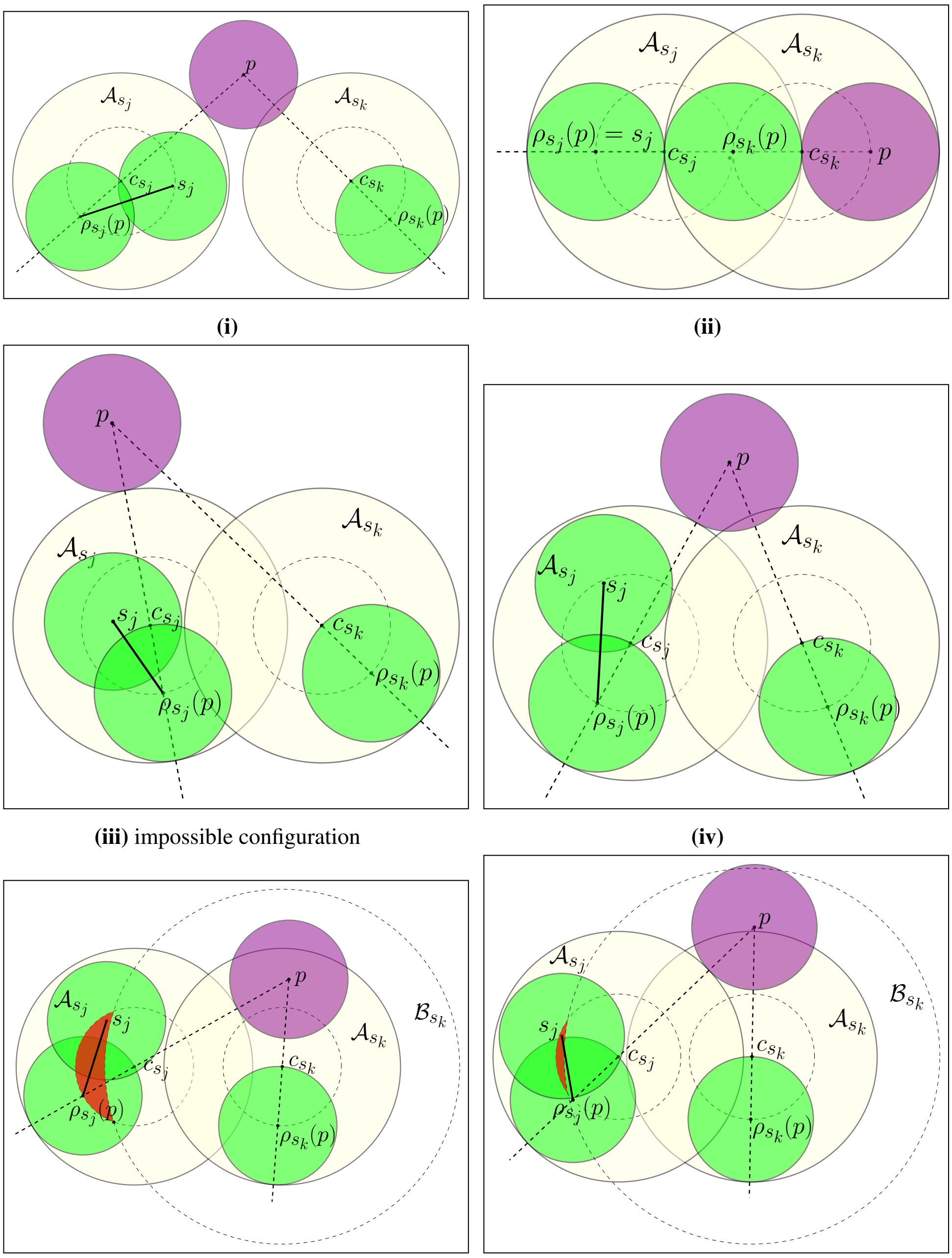

**(i)** **(ii)**

**(iii)** impossible configuration **(iv)**

**(v)** **(vi)**

**Fig. 9:** Illustration of the proof of Proposition 1.